\documentclass[onecolumn,preprint,superscriptaddress]{revtex4-2}
\usepackage{bm}
\usepackage[colorlinks=true,linkcolor=blue,citecolor=blue]{hyperref}
\usepackage{times}
\usepackage{amsmath}
\usepackage{amssymb}
\usepackage{amsthm}
\usepackage{amsfonts}
\usepackage{enumerate}
\usepackage{latexsym}
\usepackage{ifpdf}
\usepackage{natbib}
\usepackage{psfrag}
\newcommand{\beq}{\begin{equation}}
	\newcommand{\eeq}{\end{equation}}
\usepackage{graphicx}
\usepackage{makeidx}
\usepackage{color}
\usepackage{array}
\usepackage{multirow}
\usepackage{siunitx}
\usepackage{}
\usepackage{physics}
\usepackage{siunitx}

\begin{document}
	
	\title{Flux-flow instability across Berezinskii Kosterlitz Thouless phase transition in KTaO$_3$ (111) based	superconductor}

	\author {Shashank Kumar Ojha}
	\altaffiliation{Contributed equally}
	\email{shashank@iisc.ac.in}
	\affiliation{Department of Physics, Indian Institute of Science, Bengaluru 560012, India}
	\author {Prithwijit Mandal}
	\altaffiliation{Contributed equally}
	\affiliation{Department of Physics, Indian Institute of Science, Bengaluru 560012, India}
	\author {Siddharth Kumar}
	\affiliation{Department of Physics, Indian Institute of Science, Bengaluru 560012, India}
	\author {Jyotirmay Maity}
	\affiliation{Department of Physics, Indian Institute of Science, Bengaluru 560012, India}
	\author {Srimanta Middey}
	\email{smiddey@iisc.ac.in }
	\affiliation{Department of Physics, Indian Institute of Science, Bengaluru 560012, India}

	\begin{abstract}
			Abstract: The nature of energy dissipation in 2D superconductors under perpendicular magnetic field at small current excitations has been extensively studied over the past two decades. However, dissipation mechanisms at high current drives remain largely unexplored. Here we report on the distinct behavior of energy dissipation in the AlO$_\textnormal{x}$/KTaO$_3$ (111) system hosting 2D superconductivity in the intermediate disorder regime. The results show that below the Berezinskii Kosterlitz Thouless (BKT) phase transition temperature ($T_\mathrm{BKT}$), hot-spots and Larkin Ovchinnikov type flux-flow instability (FFI) are the major channels of dissipation, leading to pronounced voltage instability at large currents. Furthermore, such FFI leads to a rare observation of clockwise hysteresis in current-voltage characteristics within the temperature range $T_\mathrm{BKT} < T < T$$_{\textnormal{C}}$ ($T$$_{\textnormal{C}}$ is superconducting transition temperature). These findings deepen our understanding of how a BKT system ultimately transforms to a normal state under increasing current.
	\end{abstract}

	\maketitle
	
	\section*{Introduction}

	The ability to conduct dissipationless electrical current is one of the most striking features of a superconductor~\cite{tinkham:2004p}. The phenomena of pair breaking puts an upper theoretical bound on the maximum current that a superconductor can withstand without dissipation~\cite{bardeen:1962p667}. However,  a finite dissipation always sets in at much lower current densities in reality, leading to breakdown of the superconductivity (SC) much before the pair breaking limit is reached.  Therefore understanding dissipation mechanism is not only critical to answering some of the fundamental questions about the nature and origin of superconductivity, but will also be pivotal in realizing next generation applications such as superconducting digital memory, cavities for particle accelerators and THz radiation sources etc.~\cite{Embon:2017p85,Devoret:2013p1169,Gurevich:2008p104501,Welp:2013p9}.
	
	In 1D, phase slip centers are the primary cause of dissipation~\cite{tinkham:2004p}. In 2D, an additional complication arises due to occurrence of a topological phase transition which belongs to the Berezinskii Kosterlitz Thouless (BKT) universality class ~\cite{berezinskii:1971P493,Kosterlitz:1973P1181}.  Below the BKT phase transition temperature ($T$$_{\textnormal{BKT}}$), bound vortex-antivortex  pairs are the bare topological excitations which become unbound above the $T$$_{\textnormal{BKT}}$~\cite{Kosterlitz:1973P1181,Beasley:1979p1165,Epstein:1981p534,Resnik:1981p1542}. Nonetheless, some bound vortex-antivortex pairs still exist even in the temperature range $T$$_{\textnormal{BKT}}$ $\leq$ $T$ $\leq$$T$$_{\textnormal{C}}$ under zero electrical current  ($I$)~\cite{jos:2013p}. Application of $I$ leads to a further increase in free vortex density due to unbinding of bound vortex-antivortex pairs. These free vortices feel magnus force under the applied current and hence can move with very high velocities at large currents~\cite{Embon:2017p85}.  While the presence of ultra-fast moving vortices and its possible connection with phase slip lines (which are 2D analogue of phase slip centers) has been demonstrated earlier~\cite{Andronov:1993p193,Weber:1991p289,Sivakov:2003p267001,Paradiso:2019p025039,Berdiyorov:2009p184506}, what happens to these topological defects just before the breakdown remains puzzling. One of the proposition has been that such fast moving vortices can become unstable at large currents leading to flux-flow instability (FFI) as proposed by Larkin and Ovchinnikov (LO)~\cite{larkin1975:960,Klein:1985p413}. While such a scenario has been demonstrated under magnetic field~\cite{Klein:1985p413,Doettinger:1994p1691,samoilov:1995p4118,Ruck:1997p3378,Xiao:1998p11185,Xiao:1998pR736,Kunchur:2002p137005,Babi:2004p092510,Embon:2017p85,Dobrovolskiy:2020p3291}, its manifestation in BKT systems in absence of external magnetic field remains scarce~\cite{Saito:2020p074003}.

	The presence of disorder in samples, which is inevitable in reality, further complicates this problem by turning the BKT system inhomogeneous. Such inhomogeneities might range from atomic level point defects to macroscopically phase separated regions~\cite{Caprara:2012p196401,Caprara:2014p014002,Caprara:2013p020504,Ariando:2011p188}. While the former determines the vortex pinning strength, the latter often leads to a network of superconducting puddles joined by weak superconducting links. Such weak links, which are hosts of hot-spots, are very fragile under large electric field and are another competing source of dissipation under large current in the absence of magnetic field~\cite{Likharev:1979p101}. In the past, much of the attention has been paid to understanding the dissipation in either very clean or dirty system. Notably, all of these measurements have been primarily performed in the presence of magnetic field (under very small $I$) and very little is known about the nature of dissipation under large current~\cite{Benyamini:2019p947,Paradiso:2019p025039,Saito:2020p074003}.  Further, what happens in the intermediate disordered regime also remains an open question.

	In recent years, oxide heterostructure based interfacial superconductors have turned out to a potential platform for understanding SC in 2D limit and the focus has been primarily on  SrTiO$_3$ (STO) based systems~\cite{Reyren:2007p1196,Kozuka:2009p487,Biscaras:2010p89,Chen:2018p4008}. Recently, SC has been discovered at the interface and surface of (111) oriented KTaO$_3$ (KTO) (see Fig. \ref{fig:1}a) with $T$$_{\textnormal{C}} \sim$ 1.5-2.2 K~\cite{Changjiang:2021p716,Zheng:2021p721,Ren:2022peabn4273,Mallik:2022p4625}. The $T$$_{\textnormal{C}}$ is one order of magnitude higher than heavily investigated STO based heterostructure~\cite{Reyren:2007p1196,Pai:2018p036503} and hence has generated tremendous excitement in the field of interfacial SC. Interestingly, SC was also found to be strongly influenced by the choice of over-layer grown on KTO (111) substrate. For example, the presence of a magnetic element in the overlayer could lead to a stripe order near superconducting transition~\cite{Changjiang:2021p716}. While the current focus is on understanding the origin of higher $T_\textnormal{C}$~\cite{Liu:2023p951} and possible role of spin-orbit coupling (SOC), the nature of dissipation at large current drive remains completely unknown in KTO based systems. Surprisingly, this issue also remains unexplored for any oxide based interfacial superconductors.
	
	In this work, we investigate the underlying mechanisms that cause dissipation at high current drives in KTO (111) based interfacial superconductor. Through a combination of thorough transport measurements and analysis, we have identified strong indications of LO type FFI in association with Joule heating effects. While such a behavior had previously been observed in type II superconductors under the influence of a magnetic field~\cite{Klein:1985p413,Doettinger:1994p1691,samoilov:1995p4118,Ruck:1997p3378,Xiao:1998p11185,Xiao:1998pR736,Kunchur:2002p137005,Babi:2004p092510,Embon:2017p85,Dobrovolskiy:2020p3291}, experimental evidence of such instabilities in the absence of an external magnetic field has remained elusive until now.
	
	\section*{Results}
\begin{figure*}
	\centering{
		{~}\hspace*{0cm}
		\includegraphics[scale=.32]{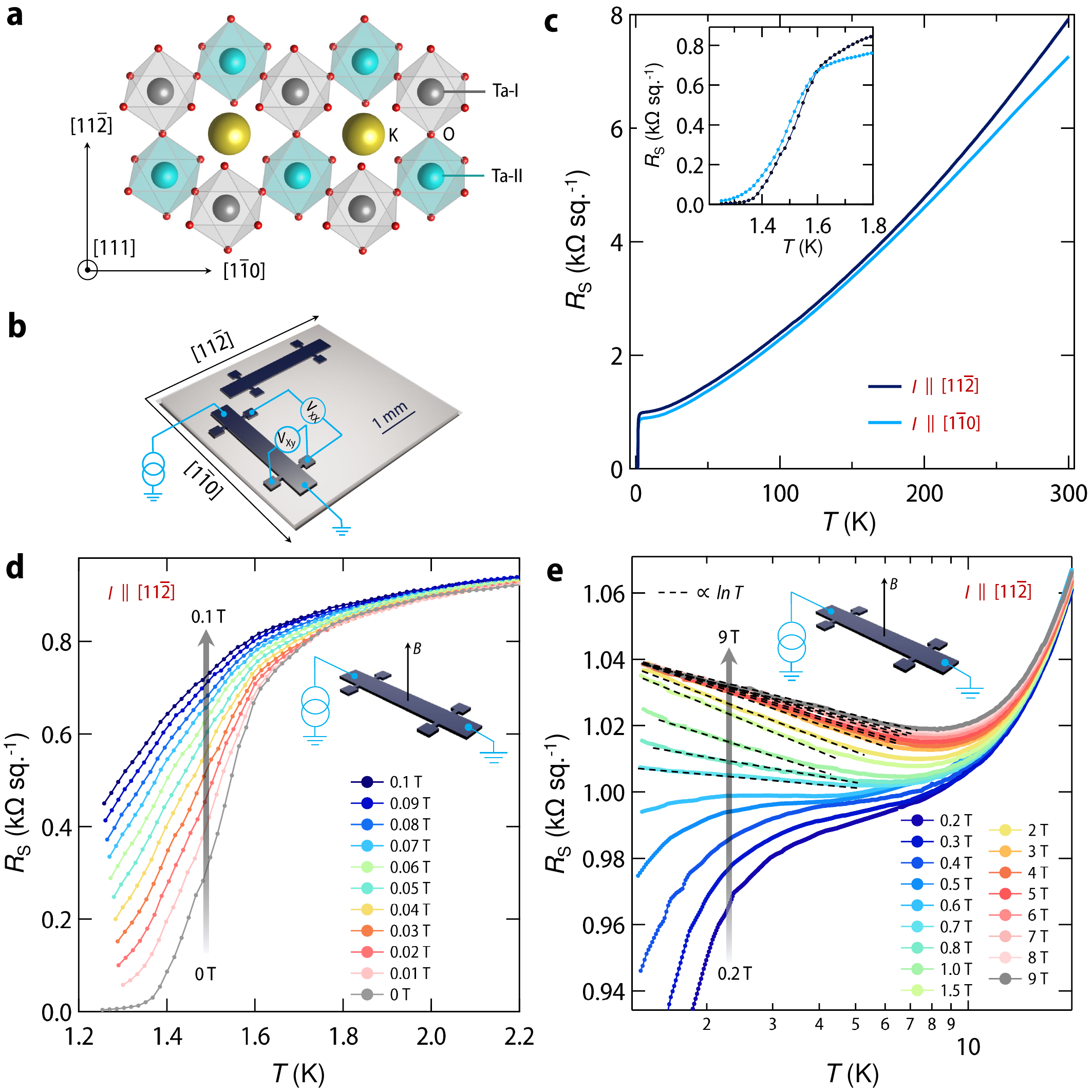}
		\caption{\textbf{Device geometry and transport behavior of AlO$_\textnormal{x}$/KTaO$_3$ (111) interface}  \textbf{a.} In a pure ionic picture, (111) oriented KTaO$_3$ can be considered as a sequence of alternating [KO$_3$]$^{5-}$ and Ta$^{5+}$ planes. Arrangement of Ta$^\text{+5}$ ions in two adjacent (111)  planes are labelled by Ta-I and Ta-II~\cite{Xiao:2011p596}.  \textbf{b.} Schematics of two Hall bars made on a  AlO$_\textnormal{x}$/KTaO$_3$ (111) heterostructure. The width of the Hall bar is $\SI{476}{\micro\meter}$ and $\SI{445}{\micro\meter}$ for [11$\bar{2}$] and [1$\bar{1}$0] directions respectively and the length between the voltage probes is 1.87 mm for both the Hall bars. \textbf{c.} Temperature-dependent $R_\textnormal{S}$ for both the Hall bars for a 7 nm AlO$_\textnormal{x}$/KTaO$_3$ (111) sample. Inset shows a magnified view around the superconducting transition temperature. The normal state  $R_\textnormal{S}$ ($T$) shows a non Fermi liquid behavior ( $R_\textnormal{S}$ $\propto$ $T$ $^\alpha$ where $\alpha$ $<$ 2) in a broad range of temperatures from 75 K to 300 K with $\alpha$ $=$ 1.5 and 1.3 for current along [11$\bar{2}$] and [1$\bar{1}$0] respectively. This behavior is in sharp contrast with the $T$ $^3$ behavior observed in bulk electron doped KTaO$_3$, where no superconductivity has been observed (see Supplementary Fig 2). Low temperature variation of  $R_\textnormal{S}$ under $B_\perp$ has been shown in \textbf{d.} (from 0 T to 0.1 T) and  \textbf{e.} (from 0.2 T to 9 T) for the Hall bar along  [11$\bar{2}$].  Dotted lines in (\textbf{e}) show logarithmic dependence of  $R_\textnormal{S}$ with the temperature near the avoided superconductor insulator transition.} \label{fig:1}}
\end{figure*}

	{\bf Two-dimensional superconductivity in AlO$_\textnormal{x}$/KTaO$_3$ (111) with intermediate disorder:} In order to avoid the potential complications caused by a magnetic overlayer on the nature of dissipation, we have fabricated a new superconducting interface by ablating non-magnetic Al$_2$O$_3$ on KTO (111) substrate [dimension 5 mm $\cross$ 5 mm $\cross$ 0.5 mm] by pulsed laser deposition technique (see Methods, Supplementary Note 1 and Supplementary Fig. 1). The resultant film is amorphous.   For electrical transport measurements,  two Hall bars were patterned along two in-equivalent crystallographic directions : [11$\bar{2}$] and [1$\bar{1}$0]  (Fig. \ref{fig:1}b)  by selective scratching of film deep into the substrate~\cite{CuiZu:2013p167}. Fig. \ref{fig:1}c shows the sheet resistance ($R_S$) vs. temperature plot of a 7 nm AlO$_\textnormal{x}$/KTO (111) sample. As evident, the interface exhibits metallic behavior down to low temperature confirming the formation of two dimensional electron gas (2DEG). The origin of the 2DEG is connected to the creation of oxygen vacancies (OVs)~\cite{Ojha:2021p054008,Mallik:2022p4625} within the top few layers of the KTO substrate. Further, a clear superconducting transition is observed with negligible anisotropy e.g. $T$$_{\textnormal{C}}$ = 1.55 K and 1.51 K for current driven along  [11$\bar{2}$] and [1$\bar{1}$0] respectively (inset of  Fig. \ref{fig:1}c) ($T$$_{\textnormal{C}}$ is estimated from the condition $R_\textnormal{S}$($T$$_{\textnormal{C}}$) = 0.5$\cross$$R_\textnormal{S}$(5 K)). While the value of $T$$_{\textnormal{C}}$ is very similar to the previous reports~\cite{Changjiang:2021p716,Zheng:2021p721}, the observation of little anisotropy is in sharp contrast with the report of large in-plane anisotropy for EuO/KTO (111) near the superconducting transition, $T$$_{\textnormal{C}}$~\cite{Changjiang:2021p716}.

	Before discussing the nature of dissipation, we first investigate the nature of this new SC system in terms of its dimensionality and the extent of the disorder. To study this, temperature dependent measurements of $R_\textnormal{S}$ ($T$) under perpendicular ($B_\perp$) and parallel ($B_\parallel$) magnetic fields have been carried out. Fig. \ref{fig:1}d shows one representative set of data for current along [11$\bar{2}$] under low $B_\perp$ (for other current orientation see Supplementary Fig. 3). Clearly, the SC is disrupted at very low magnetic field, which can be attributed to the low pinning of vortices in 2D superconductors. Upon increasing the field, the sample avoids superconductor to insulator transition around  $R_\textnormal{S}$ $\sim$ 1 k$\Omega$sq.$^{-1}$ as seen in Fig. \ref{fig:1}e. This result is in sharp contrast to the conventional theoretical framework that predicts a direct transition to an insulating state when the normal state sheet resistance approaches the quantum of resistance $h$/4e$^2$ $=$ 6.45 k$\Omega$ sq.$^{-1}$ in the limit $T$ $\rightarrow$ 0 ~\cite{Goldman:1998p39,Haviland:1989p2180}. Such a behavior is generally observed in 2D superconductors with low disorder and has proven critically important for studying phases beyond the Landau Fermi liquid theory~\cite{Kapitulnik:2019p011002}. Interestingly, at higher $B_\perp$ and lower $T$, our sample exhibits a logarithmic dependence of  $R_\textnormal{S}$ on $T$. This logarithmic divergence is incompatible with the prediction of weak localization correction in 2D or Kondo effect~\cite{Kapitulnik:2019p011002} and is connected with the emergent granular nature of our conducting interface~\cite{Beloborodov:2007p469,Zhang:2021p30}.

\begin{figure*}
	\centering{
		\hspace*{-0.35cm}
		\includegraphics[scale=.6]{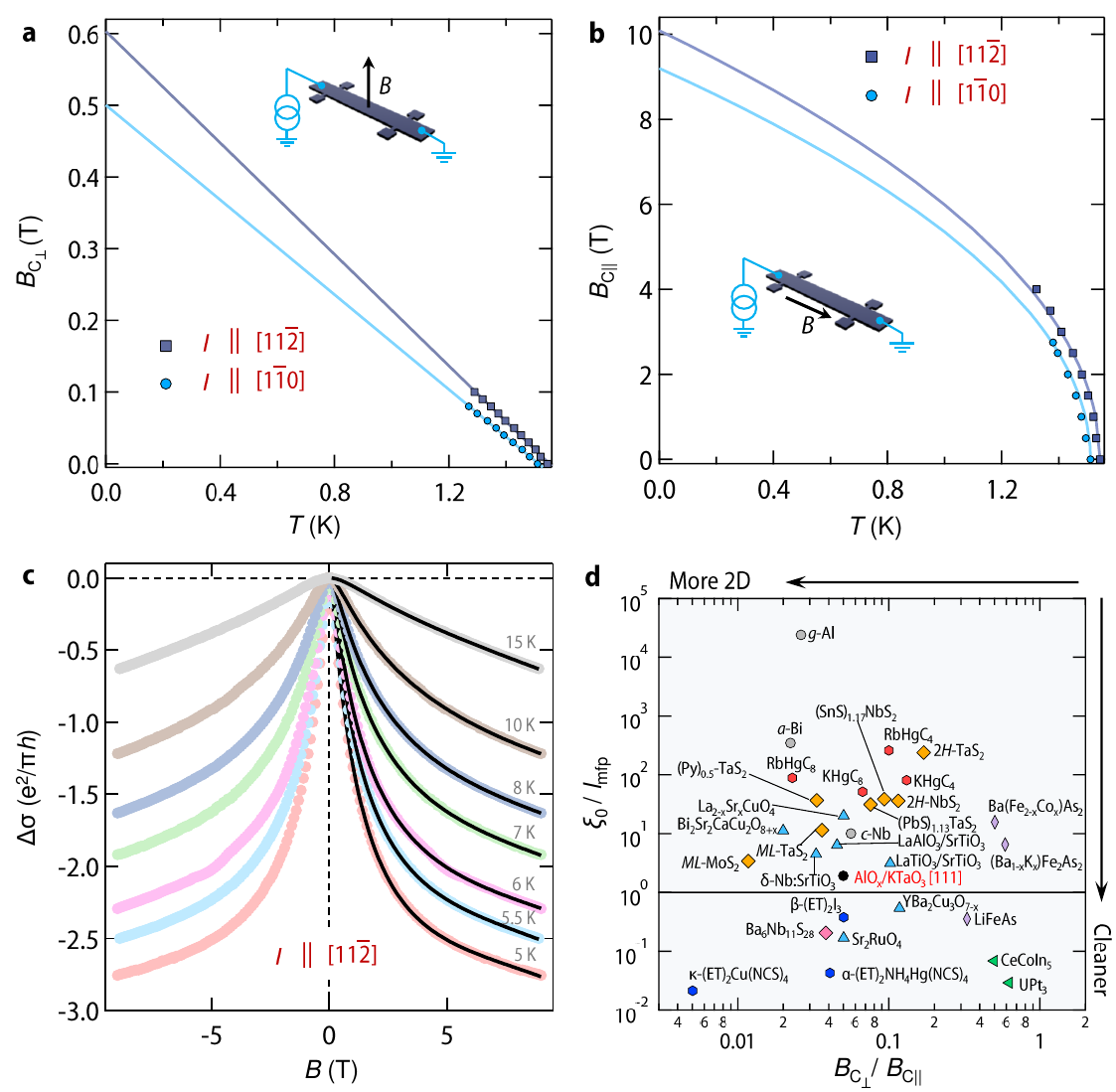}
		\caption{\textbf{Critical field, weak anti-localization, and the extent of disorder}  \textbf{a.} Temperature dependence of out of plane upper critical field ($B$$_{\text{C}}$$_\perp$) for $I$ along [11$\bar{2}$] and [1$\bar{1}$0]. The solid line denotes fitting with Ginzburg-Landau theory.  \textbf{b.} Temperature dependence of in-plane upper critical field ($B$$_{\text{C}}$$_\parallel$) for $I$ along [11$\bar{2}$] and [1$\bar{1}$0]. Further, $B$ is parallel to the current direction. The solid line denotes fitting with Tinkham’s model. \textbf{c.}  Sheet conductance difference ($\Delta$$\sigma$=$\sigma$($B$)-$\sigma$($B$=0), $\sigma$=1/$R_\textnormal{S}$ ($B$)) in the units of $\textnormal{e}$$^2$/$\pi$$h$ for the Hall bar with $I$ along [11$\bar{2}$]. The black solid curves show the fitting with ILP (Iordanskii, Lyanda-Geller, and Pikus) theory~\cite{iordanskii:1994p199,Ojha2020:p2000021} (without considering linear Rashba term) including a classical $B^2$ term (also see Supplementary Note 2 and Supplementary Fig. 7 for fitting details).
			\textbf{d.} Phase diagram of several superconducting compounds categorized by their extent of 2D character and cleanliness. 2D character is resembled by the anisotropy of critical field defined by ($B$$_{\text{C}}$$_\perp$/$B$$_{\text{C}}$$_\parallel$) and extent of disorder is quantified by the ratio between phase coherence length and electronic mean free path ($\xi_0$/$l$$_\text{mfp}$). Assuming a single isotropic band in 2D, $l$$_\text{mfp}$ is given by $l$$_\text{mfp}$=$h$/($\textnormal{e}$$^2$$k_\text{F}$$R_\textnormal{S}$), where $k_\text{F}$=(2$\pi$$n_\textnormal{s}$)$^{1/2}$ is the Fermi wave vector and $n_\textnormal{s}$ is the sheet carrier density. From the measured  $n_\textnormal{s}$ (at 5 K) and $R_S$ (at 5 K),
			the $l$$_\text{mfp}$ is estimated to be $\sim$12 nm for the present case. The value of all the parameters for other compounds have been largely taken from the reference~\cite{Devarakonda:2020p231} except for the LaTiO$_3$/SrTiO$_3$ interface which has been taken from~\cite{Biscaras:2010p89}. As evident, AlO$_\textnormal{x}$/KTaO$_3$ (111) is located very near to the  boundary between clean and dirty limits, denoted by a horizontal solid line.} \label{fig:2}}
\end{figure*}

	To verify that superconductivity is 2D in nature, out of plane and in plane upper critical fields ($B$$_{\text{C}}$$_\perp$ and $B$$_{\text{C}}$$_\parallel$) have been measured. Fig. \ref{fig:2}a shows the temperature dependence of $B_{C_\perp}$ obtained  by tracking the evolution of superconducting transition, $T$$_{\textnormal{C}}$ with  magnetic field from $R_\textnormal{S}$ vs. $T$ plots. An appreciable difference in magnitude of $B_{C_\perp}$ is observed for two configurations of current. Higher value of $B_{C_\perp}$ for current along [11$\bar{2}$] direction is consistent with the observation of higher $T$$_{\textnormal{C}}$ for current along [11$\bar{2}$] direction. The solid line shows fitting with the Ginzburg-Landau (G-L) theory which predicts a linear $T$ behavior of $B_{C_\perp}$  given by
	\begin{equation}
		B_{\textnormal{C}_\perp} =\frac {\Phi_0(1-T/T_{\textnormal{C}})}{2\pi(\xi_0)^2}
	\end{equation}
	where $\Phi_0$ is the magnetic flux quantum and $\xi_0$ is the G-L coherence length at $T$ = 0 K. $\xi_0$ from fitting is found to be $\sim$ 23.4 nm and  21.4 nm for current along [11$\bar{2}$] and [1$\bar{1}$0] directions, respectively. Fig. \ref{fig:2}b shows the temperature evolution of $B$$_{\text{C}}$$_\parallel$ for the case when current is parallel to the in plane magnetic field (see Supplementary Figs. 4, 5 and 6 for $R_\textnormal{S}$ vs. $T$ plots and data for other configurations of current). Similar to the out of plane measurement, the magnitude of $B$$_{\text{C}}$$_\parallel$ is found to be larger for the current along [11$\bar{2}$] direction. The temperature dependence of $B$$_{\text{C}}$$_\parallel$ shows a characteristic square-root dependence (shown by the solid lines in Fig. \ref{fig:2}b). Such a behavior is consistent with the Tinkham’s model~\cite{Tinkham:1963p2413} where $B$$_{\text{C}}$$_\parallel$ is given by
	\begin{equation}
		B_{\textnormal{C}_\parallel} = \frac{\Phi_0[12(1-T/T_{\textnormal{C}})]^{1/2}}{2\pi\textit{d}\xi_0}
	\end{equation}
	
	where $d$ is the effective thickness of the superconducting region. The estimated thickness of superconducting region is found to be $\sim$ 5 nm which is  much less than phase coherence length, signifying two dimensional nature of the superconductivity at the AlO$_\textnormal{x}$/KTO (111) interface.  Interestingly, the value of the in plane upper critical field extrapolated to 0 K  is found to be much larger ($\sim$10 T) than Clogston Chandrasekhar limit~\cite{Chandrasekhar:1962p7,Clogston:1962p266}. Such a large value of $B$$_{\text{C}}$$_\parallel$  is generally expected in systems with a strong SOC~\cite{Werthamer:1966p295} and  the observation of weak antilocalization characteristics in longitudinal magnetoconductance data within the normal phase (see Fig. \ref{fig:2}c) demonstrates the importance of SOC in the present case.

	In order to examine the extent of disorder in our system, we have estimated the ratio of $\xi_0$ and the electronic mean free path $l$$_\text{mfp}$. The ratio is close to 2, emphasizing that the SC at AlO$_\textnormal{x}$/KTO (111) interface falls in the intermediate disorder regime (see Fig. \ref{fig:2}d), making it an interesting system for simultaneous investigation of dissipation pertaining to an ideal BKT system and also arising from the inhomogeneous electronic structure using a single sample~\cite{Devarakonda:2020p231}. The presence of oxygen vacancies at the interface are one of the most prominent sources of disorder in the system. Clustering of oxygen vacancies can also lead to a very local inhomogeneous electronic structure in the real space~\cite{Ojha:2021p085120}. Apart from such local inhomogeneities, there is another source of inhomogeneity, which happens at a much larger scale, known as electronic phase separation (EPS)~\cite{Caprara:2012p196401,Caprara:2014p014002,Caprara:2013p020504,Ariando:2011p188}.  EPS has been routinely observed in STO based 2DEGs and is very often associated with the presence of multi carriers at the interface. The observation of two types of electrons with densities $n_1$ and $n_2$ with mobility $\mu_1$, and $\mu_2$, respectively ($n_1 >> n_2$ and $\mu_1 < \mu_2$ ) in our Hall effect measurements (see Supplementary Notes 3, 4 and 5 and accompanying Supplementary Figs. 8, 9 and 10) strongly suggests that a similar scenario can also be applicable in our samples. As a general consequence of EPS, superconducting puddles joined by weak links would emerge naturally in real space~\cite{Chen:2018p4008}, making the SC strongly inhomogeneous. This mechanism is likely a dominant cause for the observed granular nature of our system. Note that EPS could also arise due to the Rashba SOC~\cite{Caprara:2012p196401} which is also quite generic to our system.

\begin{figure*}
	\centering{
		{~}\hspace*{-0.3cm}
		\includegraphics[scale=.35]{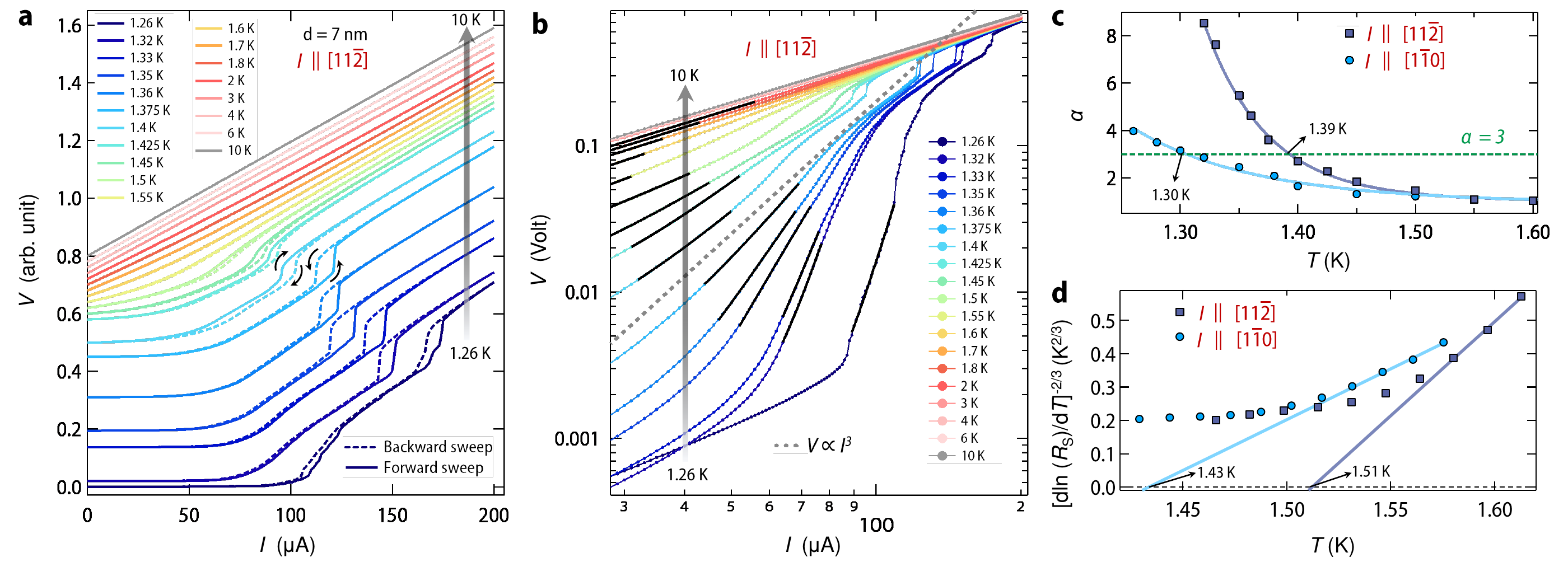}
		\caption{\textbf{Current voltage ($I$-$V$) characteristics and determination of $T$$_{\textnormal{BKT}}$}  \textbf{a.} Temperature dependent $I$-$V$ curves measured in current bias mode for the Hall bar along [11$\bar{2}$]. Solid and dotted curves denote forward and backward sweeps, respectively. Curves have been shifted upward for visual clarity. \textbf{b.} $I$-$V$ curves in logarithmic scale during the forward sweep. The solid black line shows the fit with the power law given by $V$ $\propto$ $I$$^\alpha$. A dotted gray line corresponds to $\alpha$=3 where the Berezinskii Kosterlitz Thouless transition takes place. \textbf{c.} Temperature dependence of $\alpha$ for $I$ along [11$\bar{2}$] and [1$\bar{1}$0]. A dotted green line shows a constant line for $\alpha$=3. From the crossover of $\alpha$ around 3,  $T$$_{\textnormal{BKT}}$ is found out to be 1.39 K and 1.30 K for the Hall bar along [11$\bar{2}$] and [1$\bar{1}$0], respectively  \textbf{d.} The value of $T$$_{\textnormal{BKT}}$ is also estimated using the Halperin-Nelson model ($R_\textnormal{S}$=$R_0$exp[-$b$/($T$-$T$$_{\textnormal{BKT}}$)$^{1/2}$]  where $b$ is the vortex-antivortex interaction strength)~\cite{Halperin:1979p599,Minnhagen:1987p1001}. To estimate  $T_{\textnormal{BKT}}$ using this model, dln($R_\textnormal{S}$)/d$T$]$^{-2/3}$ has been plotted as a function of $T$, near the superconducting transition temperature.  By finding the $x$ axis intercept of this plot, we find  $T$$_{\textnormal{BKT}}$ $\sim$ 1.51 K and 1.43 K for the Hall bar along [11$\bar{2}$] and [1$\bar{1}$0], respectively. These values are very close to the $T_{\textnormal{BKT}}$, obtained in (\textbf{c}).} \label{fig:3}}
\end{figure*}
	
	{\bf Various regions of dissipations as a function of $dc$ current: }
	Having established the  nature of inhomogeneities in our 2D superconducting system, we  now explore the nature of dissipation under $dc$ current bias. For this, comprehensive $I$-$V$ measurements have been performed. Fig. \ref{fig:3}a shows the $I$-$V$ curves taken in forward and backward sweeps at several fixed temperatures from 1.26 K to 10 K for current along [11$\bar{2}$] direction under zero magnetic field. All data has been shifted vertically upwards for visual clarity. Broadly four distinct regimes can be identified in the $I$-$V$ curve at the lowest temperature (1.26 K) of our measurements: (1) At small currents (less than $\SI{60}{\micro\ampere}$-$\SI{70}{\micro\ampere}$) while voltage drop looks almost independent of $I$, a small voltage drop always appears (see Fig. \ref{fig:3}b)) due to breaking of few  weakly bound vortex-antivortex pairs as the critical current for breaking of vortex-antivortex is zero~\cite{jos:2013p,Kadin:1983p6691}. (2) Above this regime, a non-linear behavior appears in a very short window from $\sim$$\SI{80}{\micro\ampere}$-$\SI{110}{\micro\ampere}$. (3)  This regime then translates into a region from $\SI{110}{\micro\ampere}$ to $\SI{175}{\micro\ampere}$, where the majority of the dissipation happens as observed by a large change in the voltage drop.  (4) Above $\SI{175}{\micro\ampere}$, the magnitude of $V$ grows almost in proportion to the applied current and finally enters into the regime of ohmic dissipation. All these different regions in $I$-$V$ characteristics are strongly $T$ dependent. The first and fourth regimes are well understood~\cite{tinkham:2004p,jos:2013p} and are skipped from further discussions.

	We first discuss the origin of non-linear $I$-$V$, observed just above the 1$^\text{st}$ regime. This regime corresponds to the intrinsic dissipation of a BKT system which is characterized by  power law behavior ($V\propto I^\alpha$) arising from current driven unbinding of thermally generated vortex-antivortex pairs near the BKT transition~\cite{Beasley:1979p1165,Epstein:1981p534}. This behavior becomes much more evident in the logarithmic plot (Fig. \ref{fig:3}b), where power law translates into a linear behavior. The value of $\alpha$ becomes exactly 3 at the $T_{\textnormal{BKT}}$ (shown by a dotted gray line ($V$ $\propto$ $I$$^3$) in Fig. \ref{fig:3}b) and is routinely used to trace out BKT phase transition in 2D superconductors.   $T_{\textnormal{BKT}}$ is estimated to be 1.39 K and 1.30 K (Fig. \ref{fig:3}c) for the Hall bar along [11$\bar{2}$] and [1$\bar{1}$0], respectively, from such analysis  (also see Fig. \ref{fig:3}d).
	
\begin{figure*}
		\centering{
			\hspace{0cm}
			\includegraphics[scale=.45]{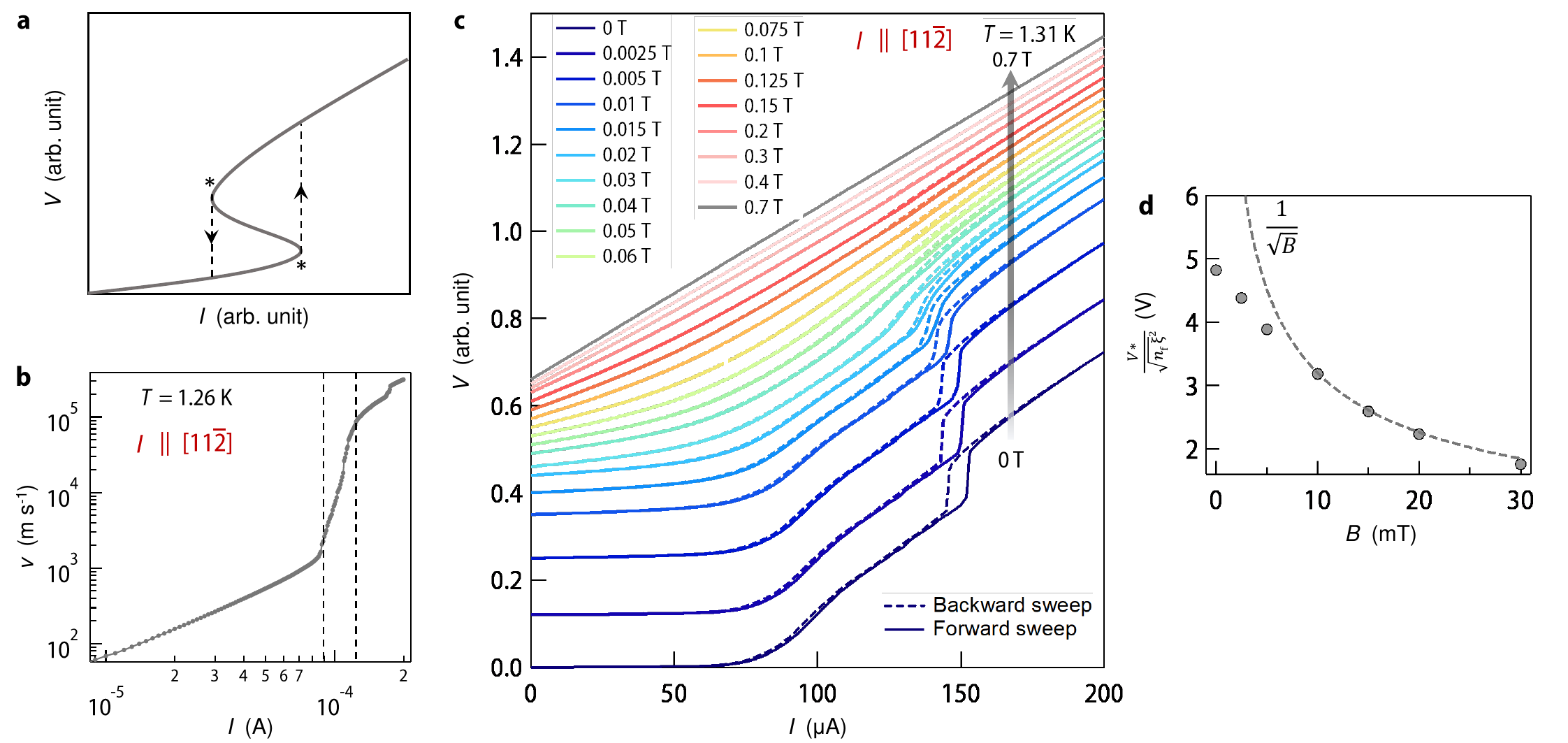}
			\hspace*{0.0cm}
			\caption{\textbf{Evidence for flux-flow instability} \textbf{a.} A schematic depicting 'S' shaped $I$-$V$ characteristics. which has two unstable points, denoted by * symbol. When the  $I$-$V$ measurement is performed in current bias mode, a voltage instability is observed when the value of current is close to unstable point leading to an abrupt increase/decrease in the voltage drop. Further, the voltage instability under backward current sweep happens at a lower current than in the forward sweep leading to a hysteresis.   \textbf{b.} Current dependent vortex velocity at 1.26 K for 7 nm AlO$_\textnormal{x}$/KTaO$_3$ (111) ($I$ along [11$\bar{2}$]). We emphasize that this whole analysis only holds only in between the region marked with the dashed lines~\cite{Saito:2020p074003}. \textbf{c.} Magnetic field dependent $I$-$V$ curves measured in current bias mode for the Hall bar along [11$\bar{2}$] on 7nm AlO$_\textnormal{x}$/KTaO$_3$ (111) sample at 1.31 K.  Curves have been shifted upward for visual clarity. \textbf{d.} Magnetic field evolution of normalized critical voltage ($\frac{V^*}{\sqrt{n_\textnormal{f}\xi^2}}$ ) calculated for Hall bar along [11$\bar{2}$] on  7 nm AlO$_\textnormal{x}$/KTaO$_3$ (111) sample. The dotted line denotes expected behavior for $\frac{1}{\sqrt{B}}$ dependence.} \label{fig:4}
		} 
	\end{figure*}

	{\bf Demonstration of  LO-type FFI: } We next focus on the nature of dissipation beyond power-law regime. At the lowest temperature of our measurement 1.26 K, which is below $T$$_{\textnormal{BKT}}$, dissipation happens via  discrete jumps in the measured voltage,  which is much more evident from the $dV$/$dI$ plot shown in Supplementary Fig. 11. These are reminiscent of phase slip events generally observed in 1D superconducting wire~\cite{tinkham:2004p}.  On the contrary,  formation of hot-spots~\cite{Gurevich:1987p941} and flux-flow instability~\cite{larkin1975:960,samoilov:1995p4118} are the two widely accepted cause for such discrete jumps under large current in thin film geometry.
		Hot-spots are the regions in real space with temperature higher than the $T$$_{\textnormal{C}}$, which appear due to the Joule self-heating in inhomogeneous systems~\cite{Skocpol:1974p4054}. In presence of hot-spots, $I$-$V$ curve takes the shape of `$S$', which would lead to a hysteresis between forward and reverse current bias (see Fig. \ref{fig:4}a). In the present case, hot-spots are most likely to occur near the weak links joining the superconducting puddles,  appearing due to the granular nature of SC as discussed earlier.

	Apart from the hot-spot effect, LO type FFI is another phenomena which leads to a `S' shape $I$-$V$ characteristics with similar voltage instabilities in current bias mode due to ultra-fast vortices~\cite{larkin1975:960,Klein:1985p413}. While the original LO instability was predicted for type II superconductors under magnetic field, we demonstrate here that such unusual phenomenon can be observed in  2D superconductors, even in absence of a magnetic field. This is due to the fact that free vortices can be generated  in 2D superconductors either by thermal fluctuation in the temperature range $T$$_{\textnormal{BKT}}$ $\leq$ $T$ $\leq$$T$$_{\textnormal{C}}$ ~\cite{Resnik:1981p1542,Doniach:p1169} or by breaking of thermally induced vortex-antivortex pairs by current below $T$$_{\textnormal{BKT}}$~\cite{Epstein:1981p534}.  In the following, we test the applicability and predictions of LO theory for the AlO$_\textnormal{x}$/KTO (111) superconductor. We further emphasize that the magnetic field generated due to current flowing through the sample or residual field in the magnet is too small and has no role for our observation (see Supplementary Notes 6 and 7 and accompanying Supplementary Figs. 12 and 13).
	
	1. The whole LO theory of FFI was built on the argument that at large vortex velocities, quasiparticles at the core of the vortex can reach energies above the superconducting energy gap ($\Delta$) due to its acceleration under electric field created due to flux-flow and ultimately diffuse away from the core. During this process, the core of the vortex starts shrinking and resultantly, the viscous damping coefficient ($\eta$) becomes a function of vortex velocity which is given by the formula~\cite{larkin1975:960,Klein:1985p413}
	
	\begin{equation}
		\eta (v)= \eta (0) \frac{1}{1+(v/v^*)^2}   \label{eq:3}
	\end{equation}
	where $v$$^*$ is the critical vortex velocity, where FFI would occur. As evident from the above equation, $\eta$ decreases with increasing $v$, leading to an ever increasing vortex velocity and after the critical velocity $v$$^*$, the system becomes unstable, leading to  a voltage jump in $I$-$V$ curve.
	
	In order to check this, we have calculated the vortex velocity (see Supplementary Note 8) using the Gor’kov–Josephson relation~\cite{Josephson:1965p419,Halperin:1979p599}. Fig. \ref{fig:4}b shows the calculated velocity  for 7 nm AlO$_\textnormal{x}$/KTO (111) sample at 1.26 K in zero magnetic field.  As evident, there is almost two orders of magnitude abrupt increase in the vortex velocity (see Fig. \ref{fig:4}b), consistent with the LO-type FFI.
		Moreover, the maximum velocity ($\sim$ 10$^5$ ms$^{-1}$) is much higher than the Abrikosov vortex velocity ($\sim$ 10$^3$ ms$^{-1}$)~\cite{tinkham:2004p}, and is also very
		similar to what has been reported earlier  for other systems  exhibiting LO-type instability under magnetic field~\cite{Dobrovolskiy:2020p3291}.
	
	2. In the original LO picture, the sample is assumed to be in perfect thermal equilibrium with the phonon bath and hence the effect of Joule heating on FFI is completely neglected. However, this may not be true in reality. In presence of overheating, a further modification has been suggested by Bezuglyj and Shklovskij~\cite{Bezuglyj:1992p234,Dobrovolskiy:2020p3291}, which would lead to a $B$ dependent $v^*$ with functional form
		\begin{equation}
			v^*\propto z \Delta^{1/2}B^{-1/2}
		\end{equation}
		where $z$ is the heat removal coefficient.
		However, this relation was derived with the constraint that  density of free vortices ($n_\textnormal{f}$) is independent of temperature, which is not the case in 2D superconductors.
		For BKT system, the expression for $v^*$ in presence of overheating can be written as (see Supplementary Note 9)
	
	\begin{equation}
		v^*\propto \Bigg(\frac{\Delta(T) \xi^2(T)}{n_\textnormal{f}(T,B)}\Bigg)^{1/2}B^{-1/2}  \label{eq:4}
	\end{equation}	
	
	\noindent Since the exact temperature dependence of $\xi$ is unknown, we rewrite the above equation using  the Gor’kov–Josephson relation~\cite{Josephson:1965p419,Halperin:1979p599} as
	
	\begin{equation}
		\frac{V^*}{\sqrt{n_\textnormal{f}\xi^2}} \propto \Delta (T)^{1/2}B^{-1/2}   \label{eq:5}
	\end{equation}
	
	\noindent where $V^*$ marks the onset of voltage instability in $I$-$V$ curve.
	
	To testify this for present case, we have performed   $I$-$V$ measurement in presence of $B$ at a fixed temperature.  Fig. \ref{fig:4}c shows one representative set of data for Hall bar along [11$\bar{2}$] on 7nm AlO$_\textnormal{x}$/KTO (111) sample. As evident from Fig. \ref{fig:4}d, normalized $V^*$ is indeed dependent on $B$ with a characteristic of $B$$^{-1/2}$ dependence at higher fields (for calculation of the denominator in eq. \ref{eq:5}, we refer to Supplementary Note 10).  We note that, a similar behavior was observed for $v^*$ in Nb-C superconductor near $T$$_{\textnormal{C}}$ and the deviation from $B$$^{-1/2}$ at low fields was  attributed to the possible role of edge controlled FFI~\cite{Dobrovolskiy:2020p3291}.
	
	\section*{Discussions}
	
	Having demonstrated the relevance of hot-spots and LO type FFI in our samples, we next discuss the temperature evolution of these two effects. For this, we first note that since the specific heat transfer power from the sample to the thermal bath is not known at a given temperature, a quantitative estimation of relative contribution from the hot-spot and FFI can not be made. Nonetheless, our temperature dependent analysis of $\frac{V^*}{\sqrt{n_\textnormal{f}\xi^2}}$  indicates that the hot-spots are most likely effective below $T_{\textnormal{BKT}}$ whereas FFI would be more applicable close to $T_{\textnormal{C}}$ (see Supplementary Note 11 and Supplementary Fig. 14 ).

\begin{figure*}
		\centering{
			\hspace{0cm}
			\includegraphics[scale=.38]{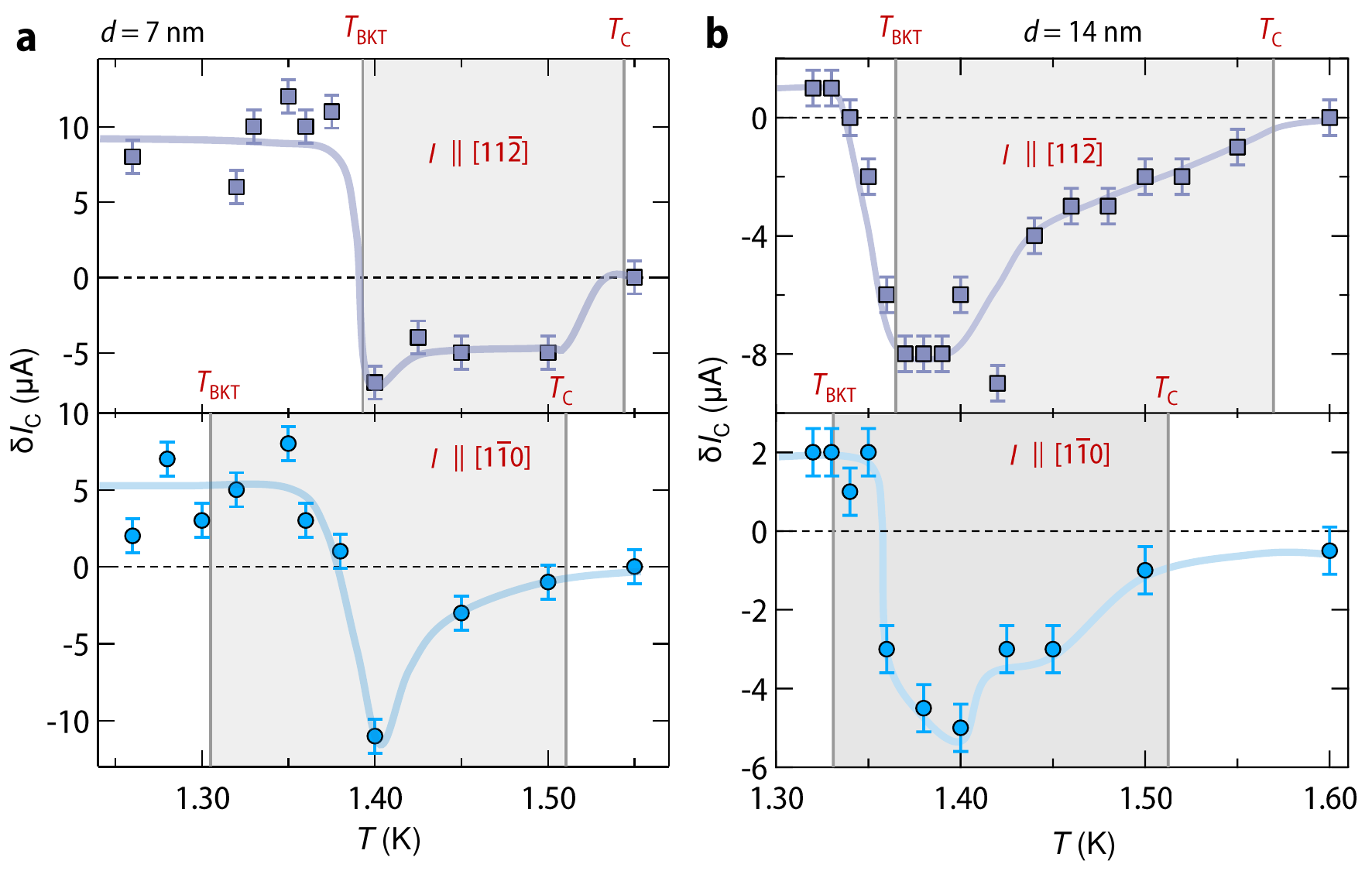}
			\hspace*{0.0cm}
			\caption{\textbf{Sign change of hysteresis across Berezinskii Kosterlitz Thouless transition} \textbf{a.} Maximum width of hysteresis  [$\delta$$I$$_\textnormal{c}$=($I$$_\textnormal{c}$)$_\text{forward}$-($I$$_\textnormal{c}$)$_\text{backward}$, ($I$$_\textnormal{c}$)$_\text{forward}$ and ($I$$_\textnormal{c}$)$_\text{backward}$ are the values of critical current in the middle of hysteresis in the forward and backward sweep, respectively] for $I$ along [11$\bar{2}$] and [1$\bar{1}$0] for 7 nm AlO$_\textnormal{x}$/KTaO$_3$ (111) sample. \textbf{b.}  Similar data for another sample with double the thickness (14 nm) of AlO$_\textnormal{x}$. The sheet carrier density for this sample was found to be 1.1$\cross$10$^{14}$cm$^{-2}$ at 300 K and the $T$$_{\textnormal{C}}$ is 1.57 K and 1.51 K for $I$ along [11$\bar{2}$] and [1$\bar{1}$0] respectively, these values are very similar to the case of 7 nm AlO$_\textnormal{x}$/KTaO$_3$ (111) sample. The change in the width of the hysteresis upon multiple cycling has been used to estimate the error bar.}
			\label{fig:5}
		}
	\end{figure*}

	We next focus on the temperature evolution of the   $I$-$V$ hysteresis [Fig. \ref{fig:3}a] in our samples.  The hysteresis is anticlockwise at the lowest temperature of our measurement, which can be attributed jointly to the formation of hot-spots and FFI as discussed earlier. Surprisingly. the nature of hysteresis changes completely from anticlockwise to clockwise above a certain temperature (highlighted by arrows in Fig. \ref{fig:3}a). We have also observed the same behavior for  another sample with 14 nm AlO$_\textnormal{x}$ thickness (see Supplementary Figs. 16, 17 and 18 for additional data on this sample). Such clockwise hysteresis is extremely rare~\cite{samoilov:1995p4118} and has never been observed in any interfacial superconductors to the best of our knowledge. To visualize this drastic change in $I$-$V$ hysteresis, we further plot the maximum width of hysteresis ($\delta$$I$$_\textnormal{c}$) as a function of temperature. Fig. \ref{fig:5}a corresponds to $\delta$$I$$_\textnormal{c}$ for Hall bar along [11$\bar{2}$] and [1$\bar{1}$0] directions (also see Supplementary Fig. 15) on 7 nm AlO$_\textnormal{x}$/KTO (111) sample. Fig. \ref{fig:5}b contains a similar set of data for the 14 nm AlO$_\textnormal{x}$/KTO (111) sample. As clearly evident, hysteresis always changes its sign  around the $T$$_{\textnormal{BKT}}$ and vanishes around $T$$_{\textnormal{C}}$ in all the four Hall bars, that we have investigated in this work. While the vanishing of anticlockwise hysteresis across $T$$_{\textnormal{BKT}}$ can be accounted by the disappearance of quasi 1D dissipating channels such as weak links~\cite{Ovadyahu:1980p375}, the clockwise hysteresis can not be explained by the hot-spot effect. Rather, the observation of clockwise hysteresis in $I$-$V$ can be explained by (i) vortex de-pinning like instabilities~\cite{Liu:2002p144510} or (ii) LO type FFI~\cite{samoilov:1995p4118}. Since our sample is already in flux-flow regime (see Supplementary Note 12 and Supplementary Fig. 19)  at currents much smaller than the current at which the voltage instability is observed,  the possibility of vortex de-pinning like instabilities can be discarded~\cite{Xiao:1998pR736}. We further recall the following proposition of Samoilov et al. in the context of LO theory~\cite{samoilov:1995p4118}. It was proposed that, once the superconductor is driven into the normal resistive state in the forward current sweep, the electron-electron (inelastic) scattering rate becomes higher (smaller $\tau$$_\textnormal{e}$)~\cite{samoilov:1995p4118}, leading to an electronic instability. This would mean that during the backward current sweep, the value of $V$$^*$ will be higher ($V$$^*$ $\sim$ $\tau$$_\textnormal{e}$$^{-1/2}$) than that of the forward sweep (see Supplementary Note 9). This would automatically move the $I$-$V$ curve towards the higher current and would result to a clockwise hysteresis, as observed here. Moreover, the vanishing of clockwise hysteresis at $T$$_{\textnormal{C}}$ is consistent with the fact that vortices do not exist above $T$$_{\textnormal{C}}$.

	\section*{Conclusions}
			
		In summary, our extensive analysis of temperature and magnetic field dependent $I$-$V$ measurement strongly emphasizes on  the definite role of heating effects and FFI in determining the nature of dissipation at large current bias in inhomogeneous BKT systems. The in-plane anisotropy observed in the onset temperature of clockwise hysteresis between the two Hall bars with $I$ along [1$\bar{1}$0] and [11$\bar{2}$] may arise from the in-plane anisotropy of critical vortex velocity for the onset of electronic instability. Such an observation is beyond the LO theory and calls for further investigations. Since the vortex structure in BKT system is strongly influenced by the presence of strong SOC~\cite{Devreese:2014p165304}, an extension of LO theory in presence of SOC and finite heating effects will be essential to understand such non-trivial feature.   Future studies will focus on measurements beyond the intermediate disorder regime under simultaneous top and bottom gate, which will provide an independent investigation  of the role of disorder and carrier density in determining the nature of dissipation under large current drive. Several recent studies, including those focused on magic angle twisted bilayer graphene~\cite{Cao:2021p7868}, MoS$_2$~\cite{Saito:2020p074003}, and NbSe$_2$~\cite{Paradiso:2019p025039}, have observed anomalies in high-current $I$-$V$ characteristics, which have been explained qualitatively in terms of vortex instability/phase-slip lines. Our findings of FFI across the BKT phase transition could serve as a framework for comprehending dissipation in such diverse class of 2D superconductors subjected to large currents. Further exploration of this highly non-equilibrium phenomenon in other systems that exhibit BKT transition, such as trapped atomic gases and neutral superfluids, would be of significant interest.
		
		\noindent\section*{Methods}
		\noindent\textbf{Sample growth and characterization:} AlO$_\textnormal{x}$/KTaO$_3$ (111) samples were fabricated by ablating a single crystalline Al$_2$O$_3$ target  on (111) oriented KTO substrate using a pulsed laser deposition system (Neocera LLC, USA) equipped with a high pressure reflection high energy electron diffraction setup (Staib instruments, Germany). A KrF excimer laser (Coherent, Germany) operated at a repetition rate of 1 Hz ($\lambda$=248 nm) and an energy density $\sim$1 Jcm$^{-2}$ (on the target) was used for ablating the target. Target to substrate distance was fixed at 5.6 cm.  The substrate was heated using a resistive heater whose temperature was maintained at 560 $^{\circ}$C during the growth. The growth chamber pressure was 5 $\cross$ 10$^{-6}$ Torr during the deposition. Immediately after the ablation, the sample was cooled to room temperature at a rate of 15$^\circ$Cmin$^{-1}$ under the vacuum. The surface morphology of the as received substrate and the film was monitored by performing atomic force microscopy (AFM) in non-contact mode using a Park AFM system. The thickness of the films was determined from X-ray reflectivity measurement performed in a lab based Rigaku Smartlab diffractometer. For more details, see Supplementary Note 1.

		\noindent\textbf{Transport Measurements:} All the transport measurements were performed in an Oxford Integra LLD system using the standard four probe method in the Hall bar geometry. Ohmic contacts were made by ultrasonically bonding Al wire. Electrical resistance was measured using a $dc$ delta mode with a Keithley 6221 current source and a Keithley 2182A nanovoltmeter and also using standard low-frequency lock-in technique. For $I$-$V$ measurements, a Keithley 2450 source meter was used in current bias mode with a sweep rate of $\SI{10}{\micro\ampere}$s$^{-1}$.

		\section*{Data availability}
		The data that support the findings of this work are available from the
		corresponding authors upon reasonable request.
		
		\vspace{0.2cm}
		

		\noindent\section*{Acknowledgements}
		
		Authors are thankful to Prof. Jak Chakhalian, Prof. Sumilan Banerjee, Prof. Manish Jain,  Prof. Vibhor Singh and Sanat Kumar Gogoi for fruitful discussions and valuable comments about the manuscript. SM acknowledges a Department of Science and Technology (DST) Nanomission grant (DST/NM/NS/2018/246),  SERB, India (Early Career Research Award: ECR/2018/001512, I.R.H.P.A Grant No. IPA/2020/000034),  MHRD, Government of India under STARS research funding (STARS/APR2019/PS/156/FS) for financial support. The authors acknowledge the AFM, XRD, and wire bonding facility at the Department of Physics, IISc Bangalore. We are also thankful to Ranjan Kumar Patel for proofreading.

		\noindent\section*{Author contribution}
		
		SM conceived and supervised the experiments. SKO, PM carried out all experiments and contributed to data analysis and interpretation. SK, JM contributed in the initial experiments.  SKO, PM, and SM wrote the paper. All authors discussed the results.

		\noindent\section*{Competing interests}
		The authors declare no competing interests.
		
		\noindent\section*{Additional information}
		The online version contains
		supplementary information.

\newpage

\setcounter{figure}{0}
\renewcommand{\thefigure}{S\arabic{figure}}

\makebox[\textwidth]{\bf \Large Supplementary Information}

\hspace{1cm}

\noindent\textbf{Supplementary Note 1: Growth and characterization of AlO$_\textnormal{x}$/KTaO$_3$ (111) sample}
\begin{figure}[h]
	\centering{
		\includegraphics[scale=0.8]{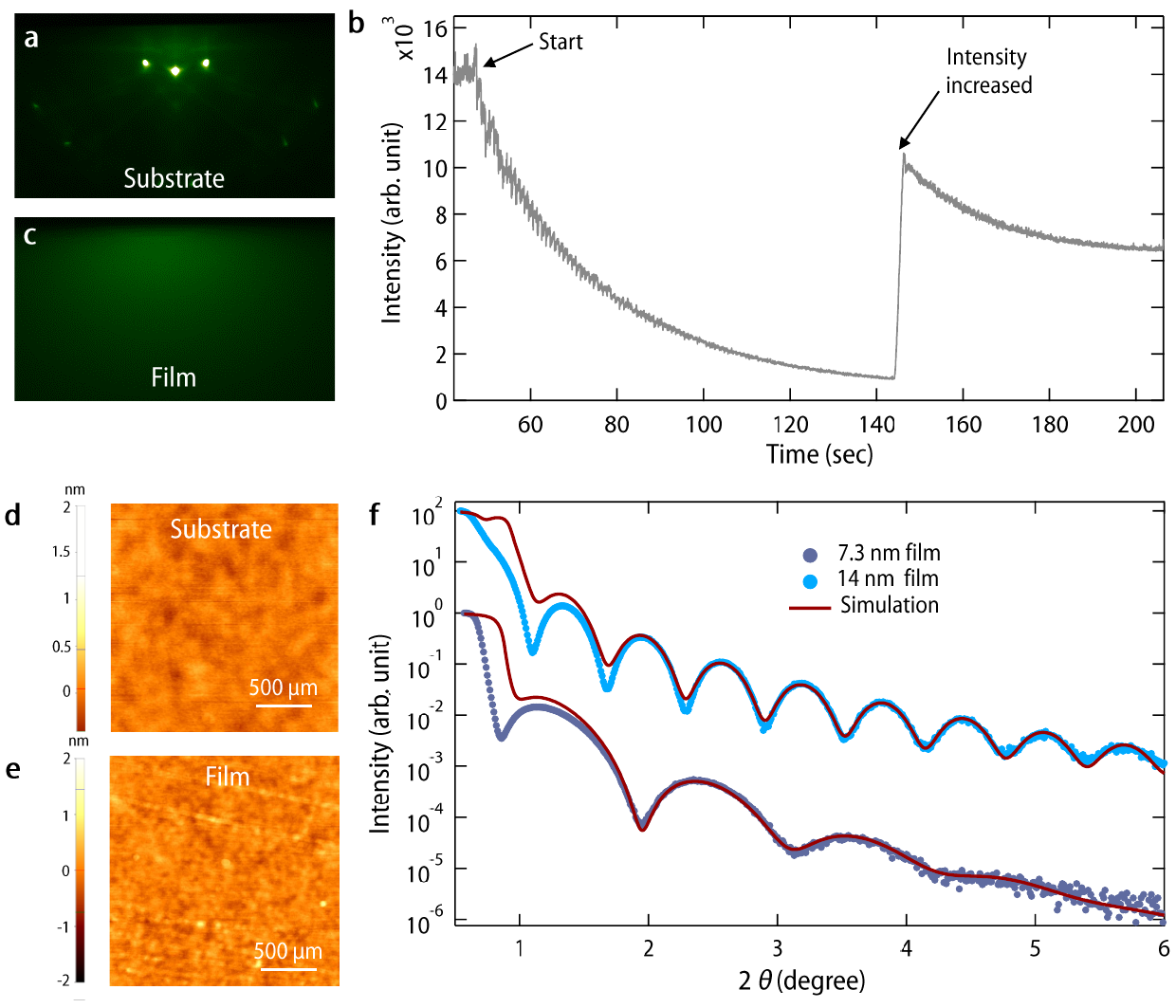}
		\caption{\textbf{a.} Reflection high energy electron diffraction image of KTaO$_3$ (111) substrate. The intensity of the specular spot during the growth of the film and just after the deposition of the film has been shown in the panels \textbf{b.} and \textbf{c.} respectively. Atomic force microscope image of the substrate and the film has been shown in panels \textbf{d.} and \textbf{e.} respectively. \textbf{f.} X-ray reflectivity pattern of the heterostructure along with the simulation. }
		\label{SFig1}}
\end{figure}

Supplementary Figure \ref{SFig1}a shows the reflection high energy electron diffraction (RHEED) image of as received KTaO$_3$ (111) substrate. Observation of intense diffraction spots along with Kikuchi lines establishes the flat and single crystalline nature of the surface and further excludes the possibility of faceting~\cite{ichimiya:2004p,Middey:2012p261602}. This is further evident from the atomic force microscopy image of the substrate, which exhibits a very smooth surface morphology with mean roughness ($R_\textnormal{q}$)$\sim$ 100 pm (Supplementary Figure \ref{SFig1}d). Supplementary Figure \ref{SFig1}b shows the temporal evolution of the intensity of the specular RHEED spot during the film deposition. As evident, intensity decreases gradually during the growth. This is due to the amorphous nature of the film which is evident from the absence of any diffraction spots in the RHEED image after the deposition (Supplementary Figure \ref{SFig1}c).  The resultant film has very flat surface morphology (Supplementary Figure \ref{SFig1}e) with $R_\textnormal{q}$$\sim$ 175 pm. In order to determine the thickness of the films, X-ray reflectivity measurements have been performed. Supplementary Figure \ref{SFig1}f shows the  measured data along with the simulation (using GenX~\cite{bjorck:2007p1174})  for two representative samples with thicknesses $\sim$7 nm and 14 nm.

\clearpage

\begin{figure}[h]
	\centering{
		\includegraphics[scale=0.7]{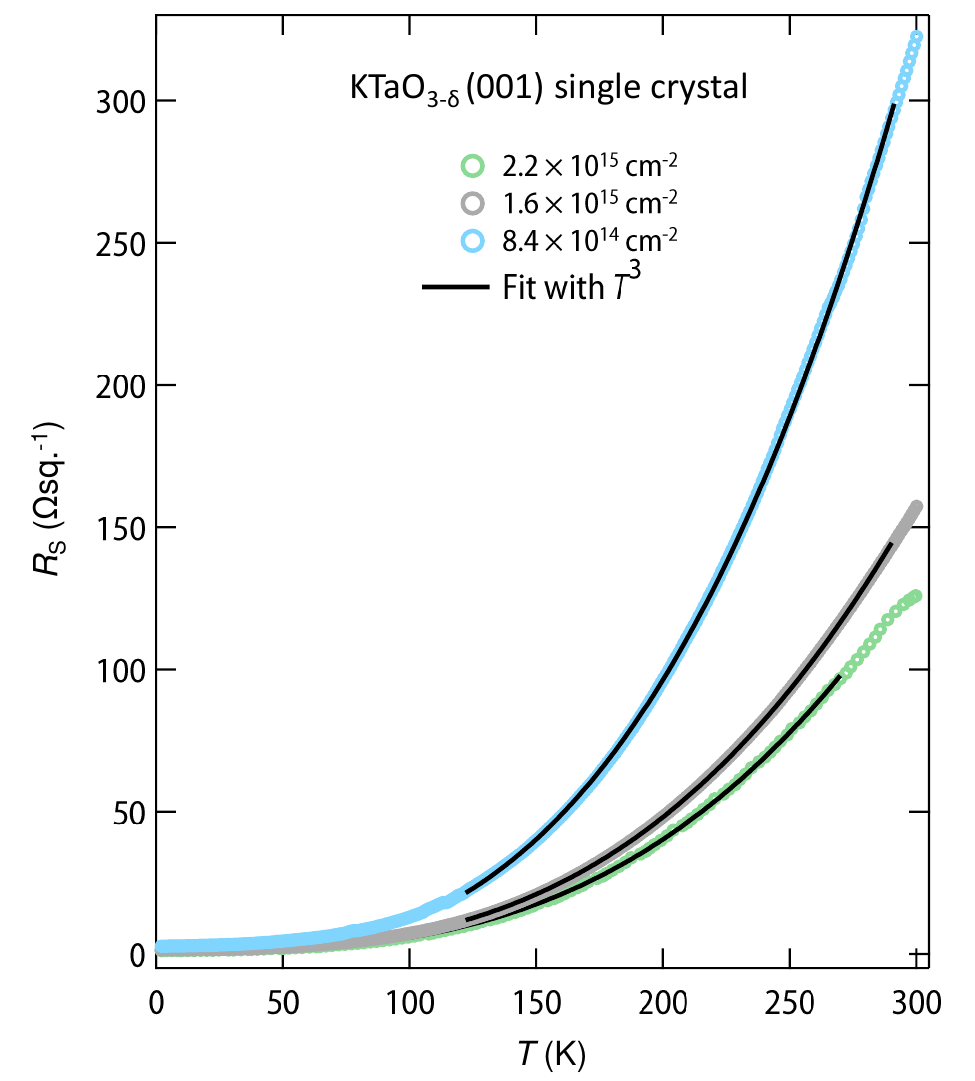}
		\caption{Sheet resistance vs. temperature plot for three bulk oxygen deficient (001) oriented KTaO$_3$ single crystal with sheet carrier densities 2.2 $\cross$ 10$^{15}$cm$^{-2}$, 1.6 $\cross$ 10$^{15}$cm$^{-2}$ and 8.4 $\cross$ 10$^{14}$cm$^{-2}$ measured at room temperature. The solid black line denotes fitting with $T$$^3$. For  more details about these samples we refer to our previous work~\cite{ojha2020:p2000021,ojha:2021p085120}.}
		\label{SFig2}}
\end{figure}

\clearpage

\begin{figure}[h]
	\centering{
		\includegraphics[scale=0.60]{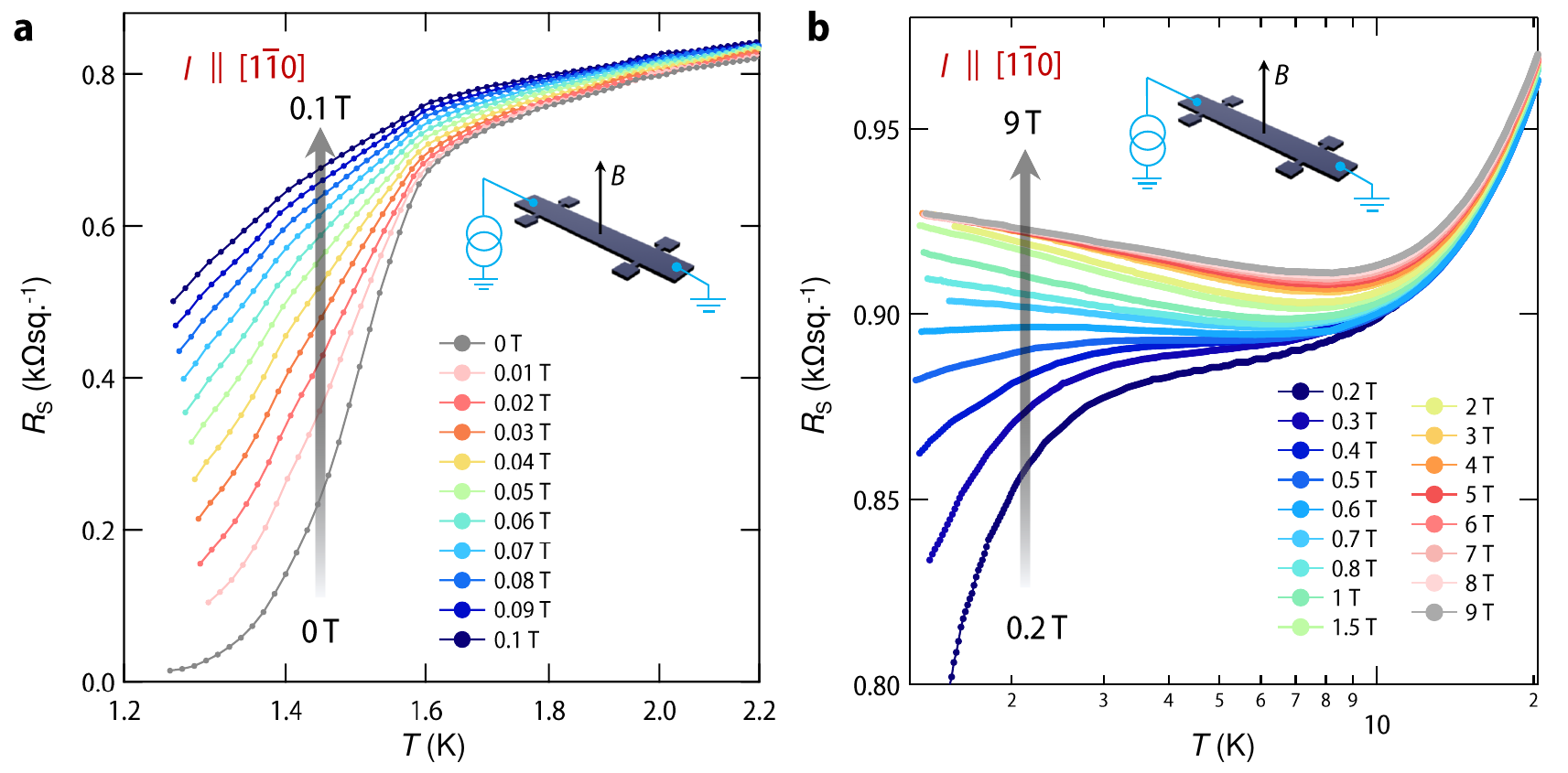}
		\caption{ Low temperature variation of $R_\textnormal{S}$ under $B_\perp$ for the Hall bar along  [1$\bar{1}$0] has been shown in \textbf{a.} (from 0 T to 0.1 T) and  \textbf{b.} (from 0.2 T to 9 T) for 7 nm AlO$_\textnormal{x}$/KTaO$_3$ (111) sample. Similar to the Hall bar along [11$\bar{2}$], an avoided superconductor insulator transition is observed around  $R_\textnormal{S}$ $\sim$ 0.9 k$\Omega$sq.$^{-1}$ A logarithmic dependence is also observed at higher $B$ and low $T$. }
		\label{SFig3}}
\end{figure}

\clearpage

\begin{figure}[h]
	\centering{
		\includegraphics[scale=0.7]{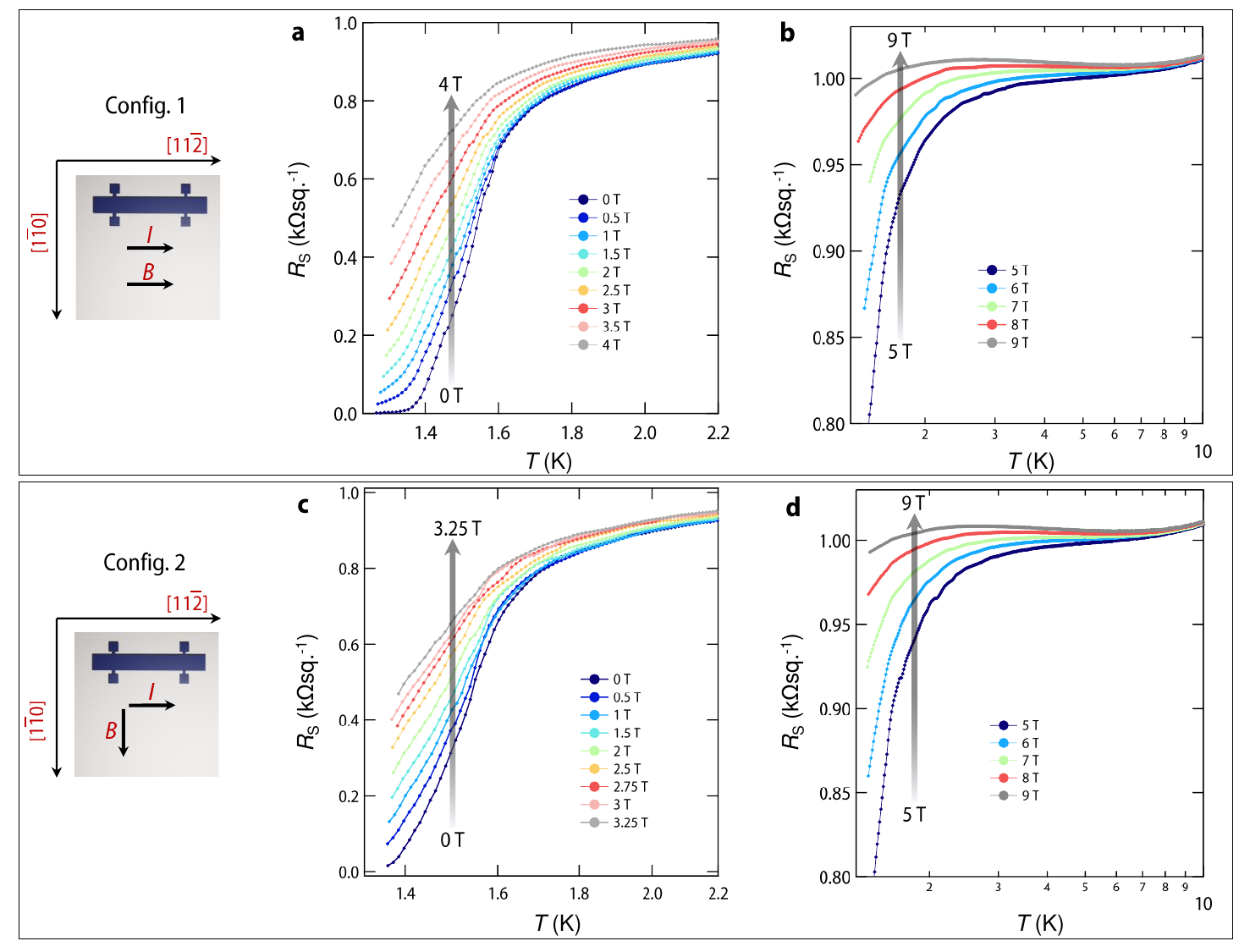}
		\caption{Low temperature variation of $R_\textnormal{S}$ under $B_\parallel$ (for 7 nm AlO$_\textnormal{x}$/KTaO$_3$ (111) sample) when $I$ $\parallel$ [11$\bar{2}$] and $B$ $\parallel$ $I$ has been shown in \textbf{a.} (from 0 T to 4 T) and  \textbf{b.} (from 5 T to 9 T). Plots for the case when $I$ $\parallel$ [11$\bar{2}$] and $B$ $\perp$ $I$ has been shown in \textbf{c.} (from 0 T to 3.25 T) and \textbf{d.} (from 5 T to 9 T).}
		\label{SFig4}}
\end{figure}
\clearpage

\begin{figure}[h]
	\centering{
		\includegraphics[scale=0.7]{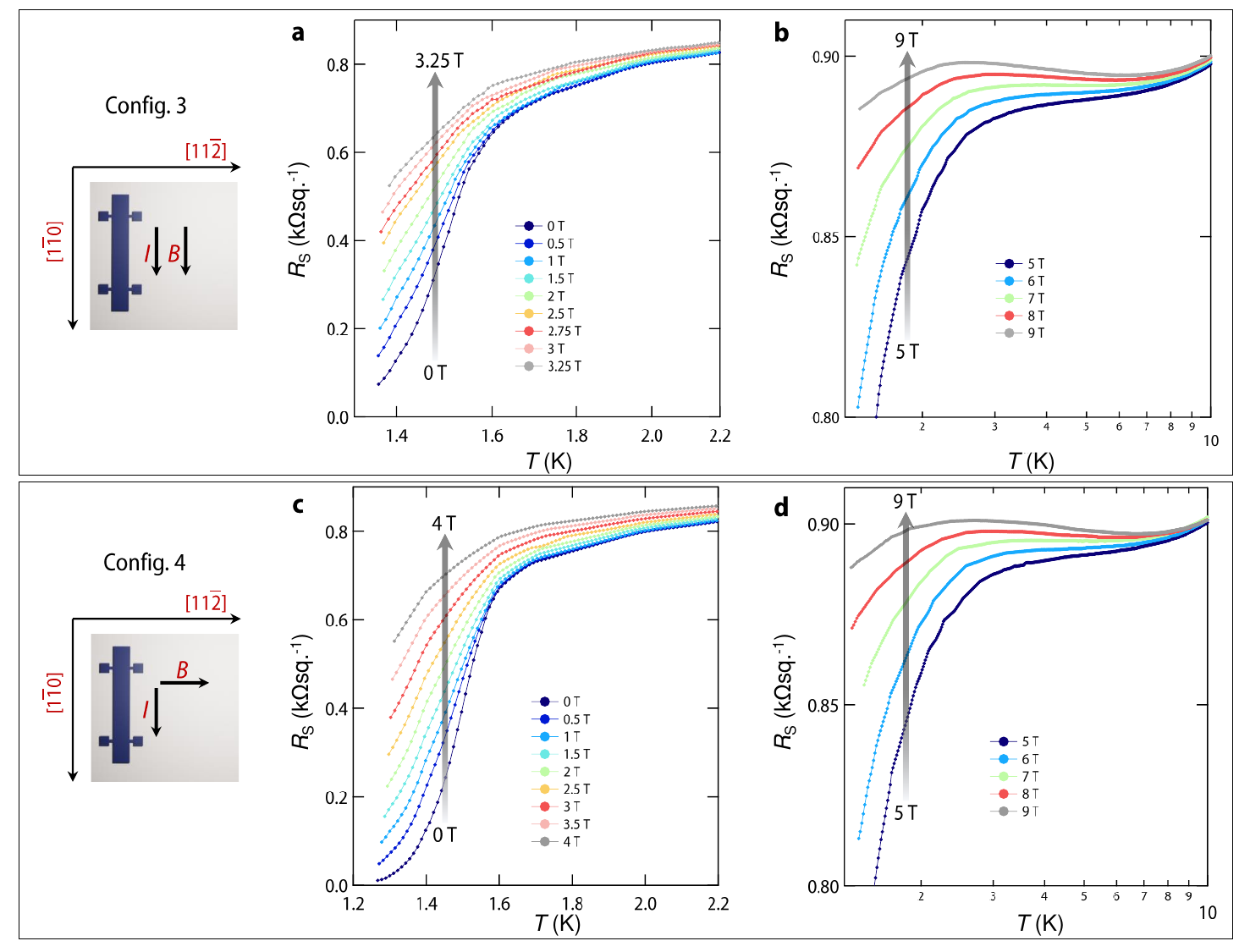}
		\caption{(a) Low temperature variation of $R_\textnormal{S}$ under $B_\parallel$ (for 7 nm AlO$_\textnormal{x}$/KTaO$_3$ (111) sample) when $I$ $\parallel$ [1$\bar{1}$0] and $B$ $\parallel$ $I$ have been shown in \textbf{a.} (from 0 T to 3.25 T) and  \textbf{b.} (from 5 T to 9 T). Plots for the case when $I$ $\parallel$ [11$\bar{2}$] and $B$ $\perp$ $I$ has been shown in \textbf{c.} (from 0 T to 4 T) and \textbf{d.} (from 5 T to 9 T).}
		\label{SFig5}}
\end{figure}
\clearpage

\begin{figure}[h]
	\centering{
		\includegraphics[scale=0.8]{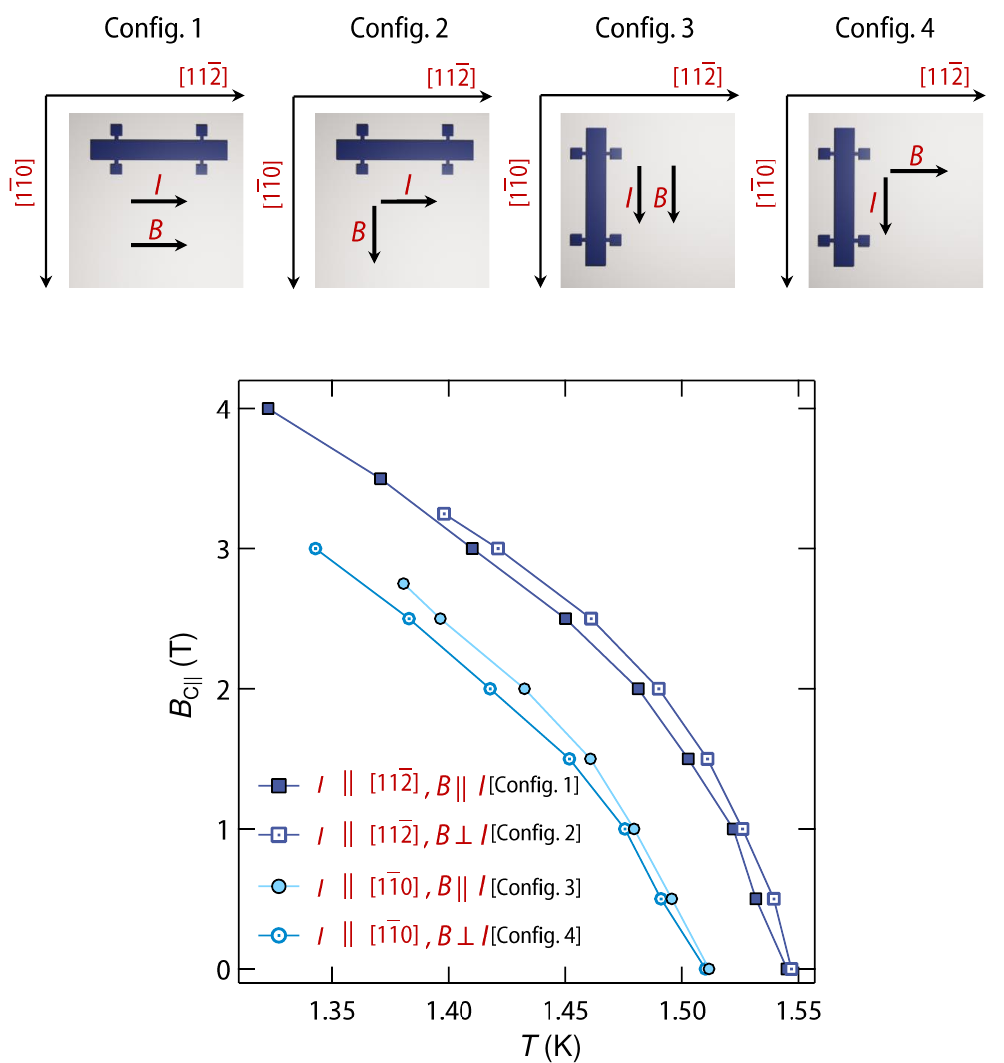}
		\caption{ Temperature dependent in-plane upper critical field ($B_{\textnormal{C}_\parallel}$), obtained  by tracking the evolution of $T$$_{\textnormal{C}}$ with  $B$$_\parallel$ in $R_\textnormal{S}$ vs. $T$ plot for all the four configurations discussed in Supplementary Figs. 5 and 6. For the case of Hall bar along [11$\bar{2}$], $B_{\textnormal{C}_\parallel}$ is found to be lower for the case when $B$ $\parallel$ $I$ (configuration 1) than $B$ $\perp$ $I$ case (configuration 2). Interestingly, this trend is completely opposite for the Hall bar along [1$\bar{1}$0] where $B_{\textnormal{C}_\parallel}$ for the case when $B$ $\parallel$ $I$ (configuration 3) is found to be higher than $B$ $\perp$ $I$ case (configuration 4). Such a behavior could arise from a small $p$-wave component in the superconducting order parameter as proposed recently~\cite{zhang:2021p}.}
		\label{SFig6}}
\end{figure}
\clearpage
\noindent\textbf{Supplementary Note 2: Fitting of weak antilocalization data with Iordanskii, Lyanda-Geller, and Pikus theory}

\begin{figure}[h]
	\centering{
		\includegraphics[scale=0.78]{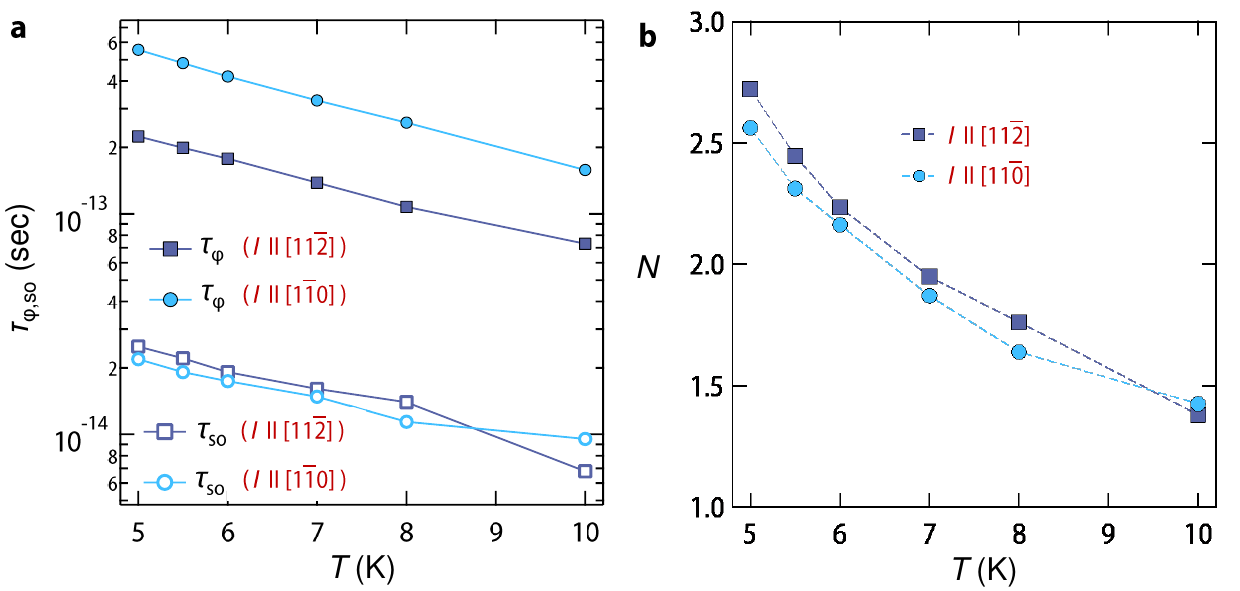}
		\caption{\textbf{a.} Temperature dependence of the phase coherence time ($\tau$$_\phi$) and spin precession time ($\tau$$_\text{SO}$) obtained from the fitting for both the Hall bars. \textbf{b.} Temperature dependence of the number of independent channels contributing to weak antilocalization. }
		\label{SFig7}}
\end{figure}

As shown in the main text, Iordanskii, Lyanda-Geller, and Pikus theory (with only cubic Rashba term) along with a small Kohler $B^2$ term provides an excellent fit at all temperatures. In the absence of linear Rashba term, the correction to the sheet conductance ($\Delta$$\sigma$) is given by

\begin{eqnarray}
\Delta\sigma (B)&=& N\frac{\textnormal{e}^2}{\pi h}\Bigg[\Psi(\frac{1}{2}+\frac{B_\phi}{B}+\frac{B_{\textnormal{SO}}}{B}) -\frac{1}{2}\Psi(\frac{1}{2}+\frac{B_\phi}{B}) \nonumber
\\ & &+\frac{1}{2}\Psi(\frac{1}{2}+\frac{B_\phi}{B}+\text{2}\frac{B_{\textnormal{SO}}}{B})-\text{ln}\frac{B_\phi+B_{\textnormal{SO}}}{B} \nonumber
\\ & & -\frac{1}{2}\text{ln}\frac{B_\phi+\text{2}B_{\textnormal{SO}}}{B}  +\frac{1}{2}\text{ln}\frac{B_\phi}{B}\Bigg] \label{seq:1}
\end{eqnarray}

where $N$ is the number of independent interference channels~\cite{Nakamura:2020p1161}, $\Psi$ is the digamma function, $B_\phi$=$\frac{\hbar}{\text{4e}l_{\phi}^2}$ ($l_{\phi}$ is the phase coherence length) and $B_{\textnormal{SO}}$=$\frac{\hbar}{\text{4e}l_{\textnormal{SO}}^2}$ ($l_{\textnormal{SO}}$ is the spin-precession length) where $\hbar$ is the reduced Planck's constant. Associated characteristics time scales, phase coherence time ($\tau$$_\phi$) and spin precession time ($\tau$$_\text{SO}$) are given by $\tau$$_\phi$=$\frac{\hbar}{\text{4e}DB_\phi}$ and $\tau$$_\text{SO}$=$\frac{\hbar}{\text{4e}DB_\text{SO}}$  where $D$ is the diffusion coefficient given by $D$=$v_\textnormal{f}$$^2$$\tau$/2 ($v_\textnormal{f}$ is the Fermi velocity and $\tau$ is the elastic scattering time). $v_\textnormal{f}$ and $\tau$  were estimated assuming a single band having parabolic dispersion with effective mass $m$$^*$= 0.3$m$$_\textnormal{e}$~\cite{Bareille:2014p3586}. Supplementary Figure \ref{SFig7}a shows the  temperature dependent  $\tau$$_\phi$ and $\tau$$_\text{SO}$ obtained from the fitting for both the Hall bars along [$11\bar{2}$] and [$1\bar{1}0$]. As evident, $\tau$$_\text{SO}$ is smaller than $\tau$$_\phi$ satisfying the criteria for weak-anti localization~\cite{BERGMANN:1984p1,Caviglia:2010p126803}. Temperature dependence of the number of independent channels  has been plotted in Supplementary Figure \ref{SFig7}b.

\clearpage

\noindent\textbf{Supplementary Note 3: Transverse resistance ($R_\textnormal{xy}$) as a function of the out-of-plane magnetic field at different temperatures}

\begin{figure}[h]
	\centering{
		\includegraphics[scale=0.8]{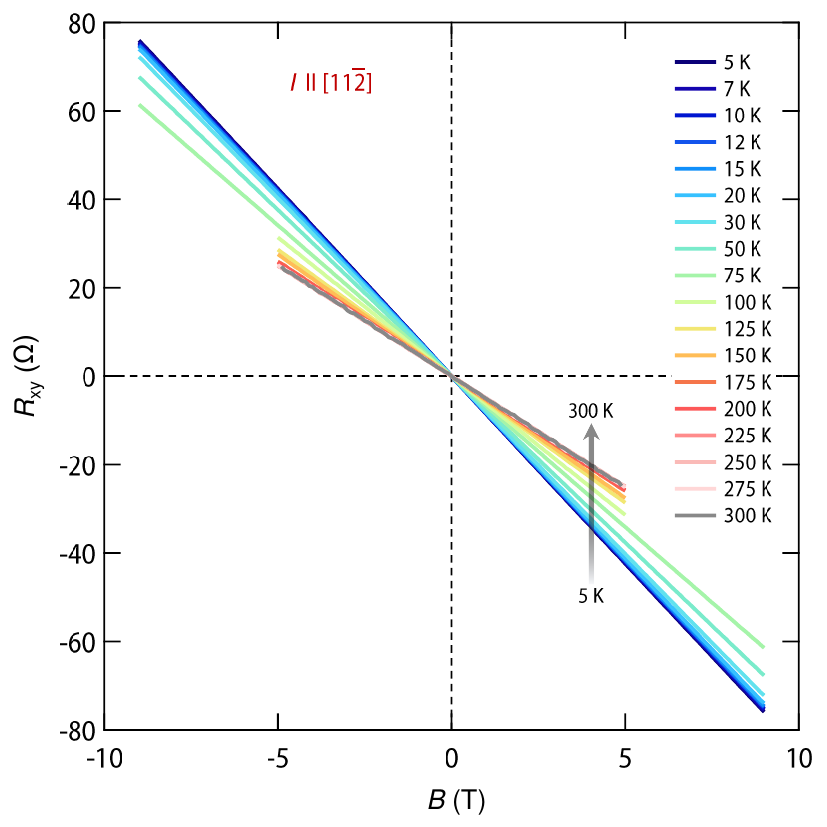}
		\caption{ Antisymmetrized transverse resistance as  a function of the magnetic field measured at various temperatures in the metallic phase for 7 nm AlO$_\textnormal{x}$/KTaO$_3$ (111) sample. The direction of $I$ is along the crystallographic axis [$11\bar{2}$].}
		\label{SFig8}}
\end{figure}

Supplementary Figure \ref{SFig8} shows $R_\textnormal{xy}$ vs. $B$ curves recorded at several fixed temperatures. To eliminate the longitudinal component of resistance ($R_\textnormal{xx}$), the data has been antisymmetrized with respect to $B$. As clearly evident, the slope $R_\textnormal{xy}/B$ increases with the decrease in temperature. Assuming single band transport, the Hall coefficient is given by $R_\textnormal{H}$ = -1/$n$e (where $n$ is the carrier density and -e is the electron's charge). Since, $R_\textnormal{xy}/B$ =$R_\textnormal{H}$, an increase in slope $R_\textnormal{xy}/B$ with lowering of temperature immediately suggests decreasing $n$ with lowering of temperature. A similar trend has also been observed for other Hall bar along [$1\bar{1}$0]. In Supplementary Note 4 and 5, we discuss the origin of the nonlinear Hall effect in the present case.

\clearpage
\noindent\textbf{Supplementary Note 4: Non-linear Hall effect and evidence for two-band transport}

\begin{figure}[h]
	\centering{
		\includegraphics[scale=0.75]{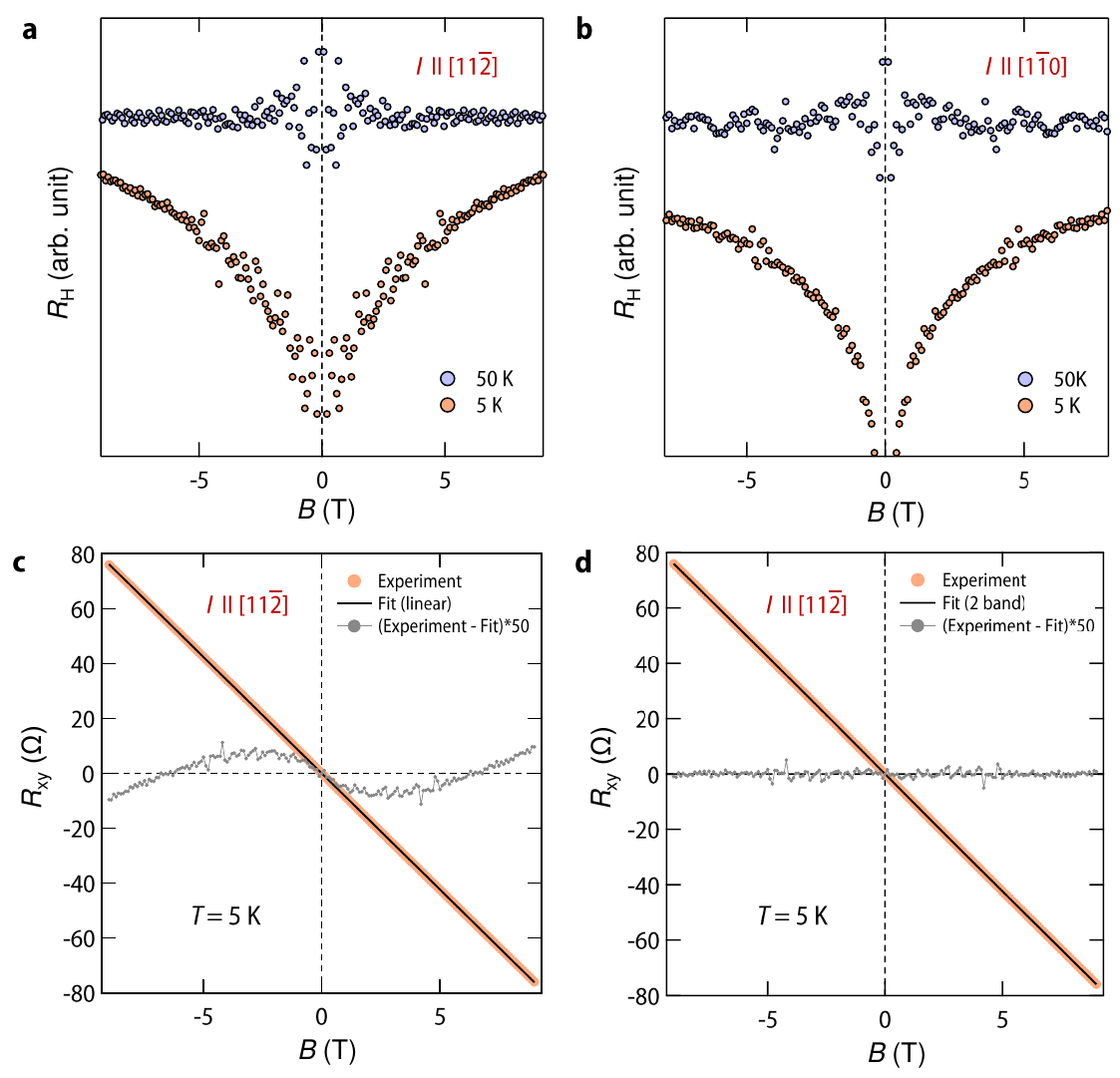}
		\caption{The Hall coefficient $R_\textnormal{H}$ as a function of out-of-plane magnetic field at 5 K and 50 K for both the Hall bars along [$11\bar{2}$] and [$1\bar{1}$0] has been shown in \textbf{a} and \textbf{b}. \textbf{c.} Linear fitting of $R_\textnormal{xy}$ vs. $B$ data at 5 K for $I$ along [$11\bar{2}$]. To visualize the presence of little non-linearity in $R_\textnormal{xy}$, the residual ($R_\textnormal{xy}$-Fit) is multiplied by 50. \textbf{d.} $R_\textnormal{xy}$ vs. $B$ data at 5 K (for $I$ along [$11\bar{2}$]) along with fitting using two band model.}
		\label{SFig9}}
\end{figure}

To extract the value of $n$, we have tried to fit the data with a straight line assuming one band model.  Interestingly, we find that, below 20 K, $R_\textnormal{xy}$ vs. $B$ can not be captured using one band approximation due to the presence of a little non-linearity. This is clearly evident in the $R_\textnormal{H}$ vs. $B$ plot shown in Supplementary Figure \ref{SFig9}a and Supplementary Figure \ref{SFig9}b for both the Hall bars. Supplementary Figure \ref{SFig9}c shows the failure of one band model in describing our Hall data at 5 K. Such nonlinear effects in $R_\textnormal{xy}$ could arise from the presence of multi carrier transport at the interface. In order to verify this, we have considered a minimal two band model where $R_\textnormal{xy}$ is given by

\begin{equation}
R_\textnormal{xy} = - \frac{1}{\textnormal{e}}\frac{(\frac{n_1\mu_1^2}{1+\mu_1^2B^2}+\frac{n_2\mu_2^2}{1+\mu_2^2B^2})B}{(\frac{n_1\mu_1}{1+\mu_1^2B^2}+\frac{n_2\mu_2}{1+\mu_2^2B^2})^2+(\frac{n_1\mu_1^2}{1+\mu_1^2B^2}+\frac{n_2\mu_2^2}{1+\mu_2^2B^2})^2B^2}       \label{seq:2}
\end{equation}

with the constraint $(\textnormal{e}.R_\textnormal{S})^{-1}$ = $n_1\mu_1 + n_2\mu_2$. Here, $n_1$, $n_2$ and $\mu$$_1$, $\mu$$_2$ are the sheet carrier densities and mobilities of the two types of electrons. As evident from the Supplementary Figure \ref{SFig9}d, two band model provides an excellent fit to our Hall data in the whole range of $B$ strongly indicating the presence of two types of carriers in the system. Such a two band transport has not been demonstrated so far for KTaO$_3$ (111) based superconductors.

\clearpage
\noindent\textbf{Supplementary Note 5: Temperature dependent sheet carrier density ($n_\textnormal{S}$), mobility ($\mu$) and evidence for carrier freezing effect}

\begin{figure}[h]
	\centering{
		\includegraphics[scale=0.55]{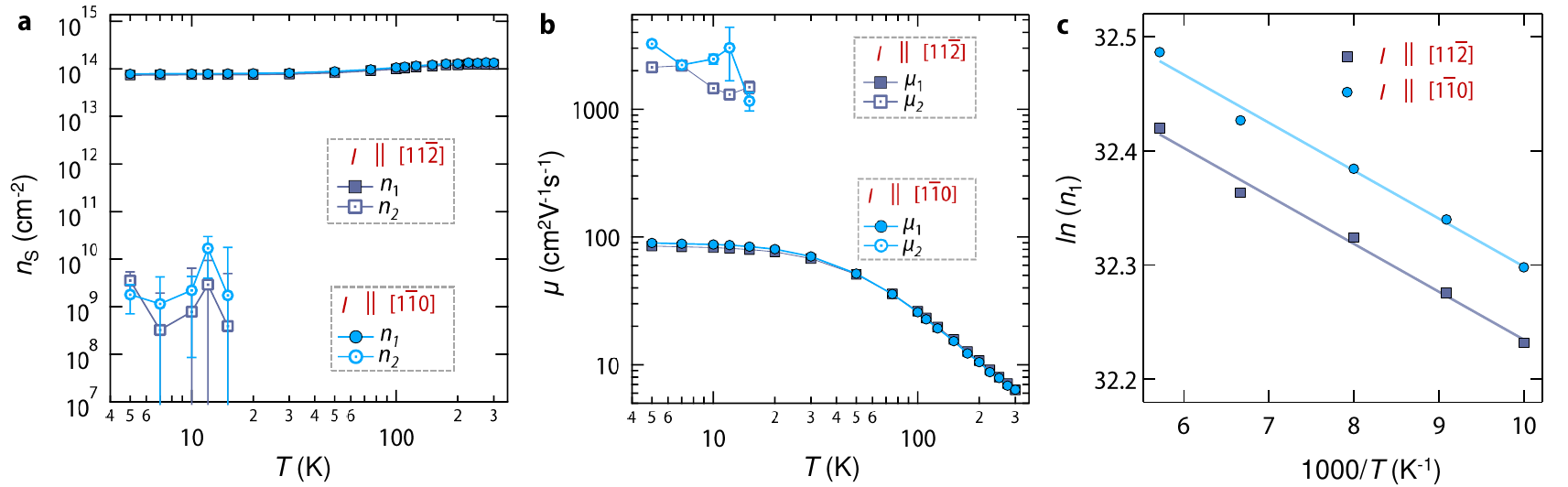}
		\caption{ \textbf{a.} Temperature dependent  $n_\textnormal{S}$ (for Hall bar along [11$\bar{2}$] and [1$\bar{1}$0])  obtained from fitting of antisymmetrized $R$$_\text{xy}$. $n_1$ and $n_2$ are the density of electrons confined to the lower and upper band, respectively.  \textbf{b.} Temperature dependent  mobility ($\mu$) for Hall bar along [11$\bar{2}$] and [1$\bar{1}$0]. $\mu_1$ and $\mu_2$ are electron's mobility confined to the lower and upper band, respectively. \textbf{c.} The Arrhenius plot of $ln$ ($n_1$) for the temperature range 100 K–175 K for both the Hall bars.}
		\label{SFig10}}
\end{figure}

Supplementary Figure \ref{SFig10}a shows temperature dependent $n_1$ and $n_2$ obtained for $I$ along [11$\bar{2}$] and [1$\bar{1}$0]. Surprisingly, $n_2$ is found out to be $\sim$10$^9$cm$^{-2}$ which is 5 orders of magnitude lower than  $n_1$.  Supplementary Figure \ref{SFig10}b shows the corresponding variation of mobility. The mobility of low density carriers ($\mu_2$) is found to be higher than that of high density carriers ($\mu_1$). Interestingly, a prominent carrier freezing effect~\cite{Liu:2011p146802} is observed below 175 K down to 100 K. This is evident from the Arrhenius plot of $ln$ ($n_1$) vs. 1000/$T$ shown in the Supplementary Figure \ref{SFig10}c. A linear fit results in a very shallow defect state, which would be just 3.6 meV below the conduction band.
\clearpage

\begin{figure}[h]
	\centering{
		\includegraphics[scale=0.6]{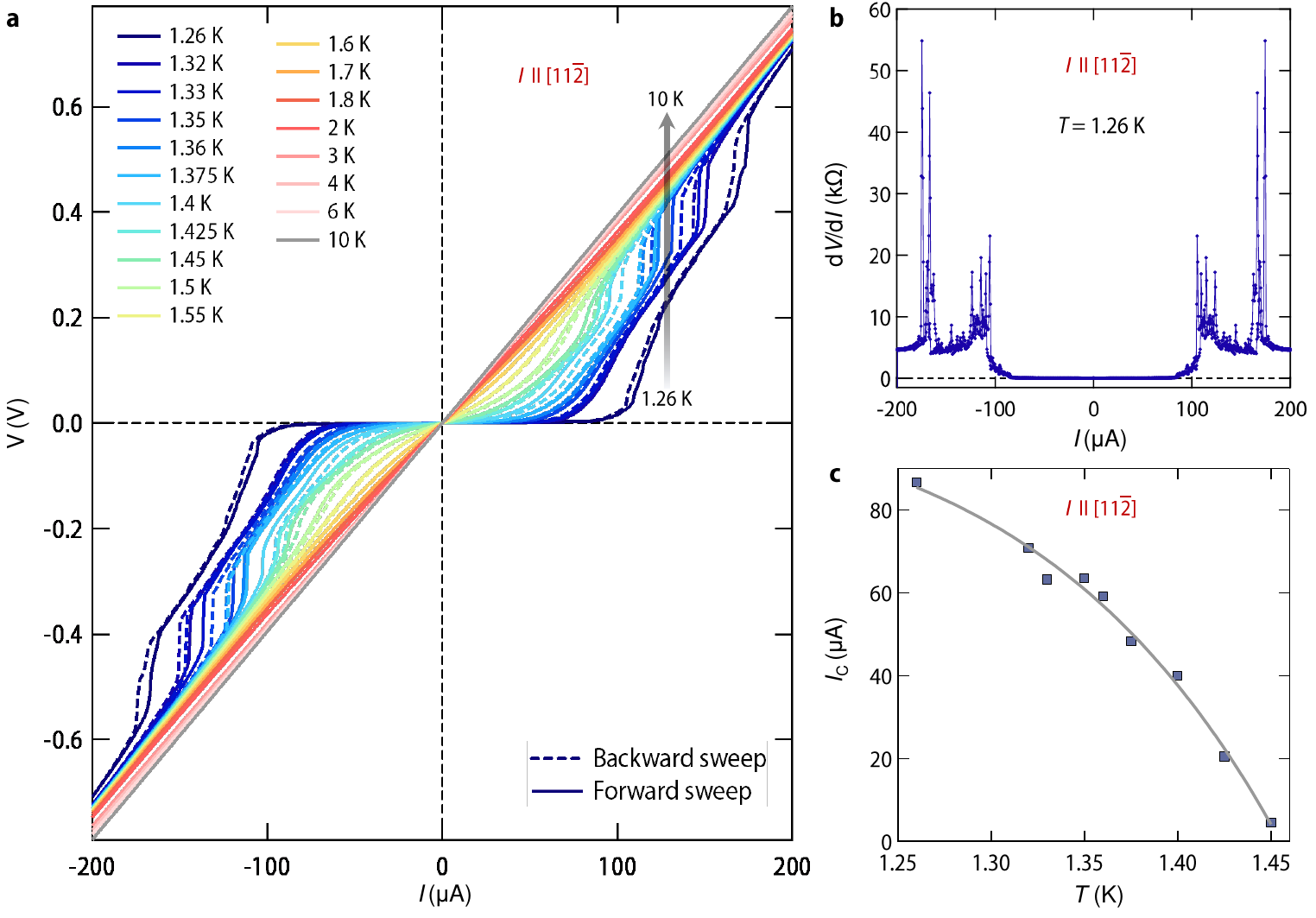}
		\caption{\textbf{a.} Temperature dependent full cycle  $I$-$V$ curves measured in current bias mode for the Hall bar along [11$\bar{2}$] on 7 nm AlO$_\textnormal{x}$/KTaO$_3$ (111) sample. For full cycle measurement, the current was swept from $\SI{0}{\micro\ampere}$$\rightarrow$$\SI{200}{\micro\ampere}$$\rightarrow$-$\SI{200}{\micro\ampere}$$\rightarrow$$\SI{200}{\micro\ampere}$$\rightarrow$$\SI{0}{\micro\ampere}$. For the sake of clarity, $\SI{0}{\micro\ampere}$$\rightarrow$$\SI{200}{\micro\ampere}$ and $\SI{200}{\micro\ampere}$$\rightarrow$$\SI{0}{\micro\ampere}$ branches have not been shown in the plot. $\SI{200}{\micro\ampere}$$\rightarrow$-$\SI{200}{\micro\ampere}$ branch is denoted as backward sweep and -$\SI{200}{\micro\ampere}$$\rightarrow$$\SI{200}{\micro\ampere}$ branch is denoted as forward sweep. \textbf{b.} $dV$/$dI$ plot at 1.26 K for $I$ along [11$\bar{2}$]. Several spikes in the derivative above certain current correspond to several discrete jumps in the voltage drop. \textbf{d.} Temperature dependence of critical current ($I_\textnormal{C}$).}
		\label{SFig11}}
\end{figure}

\clearpage

\noindent\textbf{Supplementary Note 6: Effect of self-field generated due to applied current in the sample}

In order to check the impact of magnetic field generated due to the applied current in the sample we have estimated it's order of magnitude following the results described in the reference~\cite{BabaeiBrojeny:2005p888}. Since the in-plane component of the magnetic field would have null effect on the vortices, we have only estimated the out of plane component. In the Supplementary Figure \ref{SFig12}b we have plotted the variation of the out of plane component of the magnetic field ($B_\textnormal{y}$) across the sample width within the superconducting strip (for sample geometry see Supplementary Figure \ref{SFig12}a). As evident, $B_\textnormal{y}$ peaks at edges of the sample boundary and has the maximum value of $\sim$ 3$\mu$$_0$$b$$J_\textnormal{p}$/$\pi$, where $\mu$$_0$ is the permeability of vacuum and $J_\textnormal{p}$ is pair breaking current density. Using the parameters for our sample, the obtained value of ($B_\textnormal{y}$)$_{max}$ at 1.26 K for 7 nm AlO$_\textnormal{x}$/KTaO$_3$ (111) sample for Hall bar along [11$\bar{2}$]  would be $\sim$ 10$^{-8}$ Tesla (for the sake of simplicity we have taken the value of pair breaking current just after the voltage instability). 
We now move one step further to back calculate the critical vortex velocity required to produce a LO type voltage instability with a magnetic field of $\sim$ 10$^{-8}$ Tesla by using the formula $V^*$= $v^*$$B$$L$ where $L$ is the length between the voltage probes. Putting the value of $V^*$ and $L$ yields a gigantically large value of $v^*$ $\sim$ 10$^{10}$ ms$^{-1}$. This is two orders of magnitude higher than the speed of the light and hence is unphysical, emphasizing that self-field generated due to the applied current in the system would be insufficient to lead to LO type instability.

\begin{figure}[h]
	\centering{
		\includegraphics[scale=0.52]{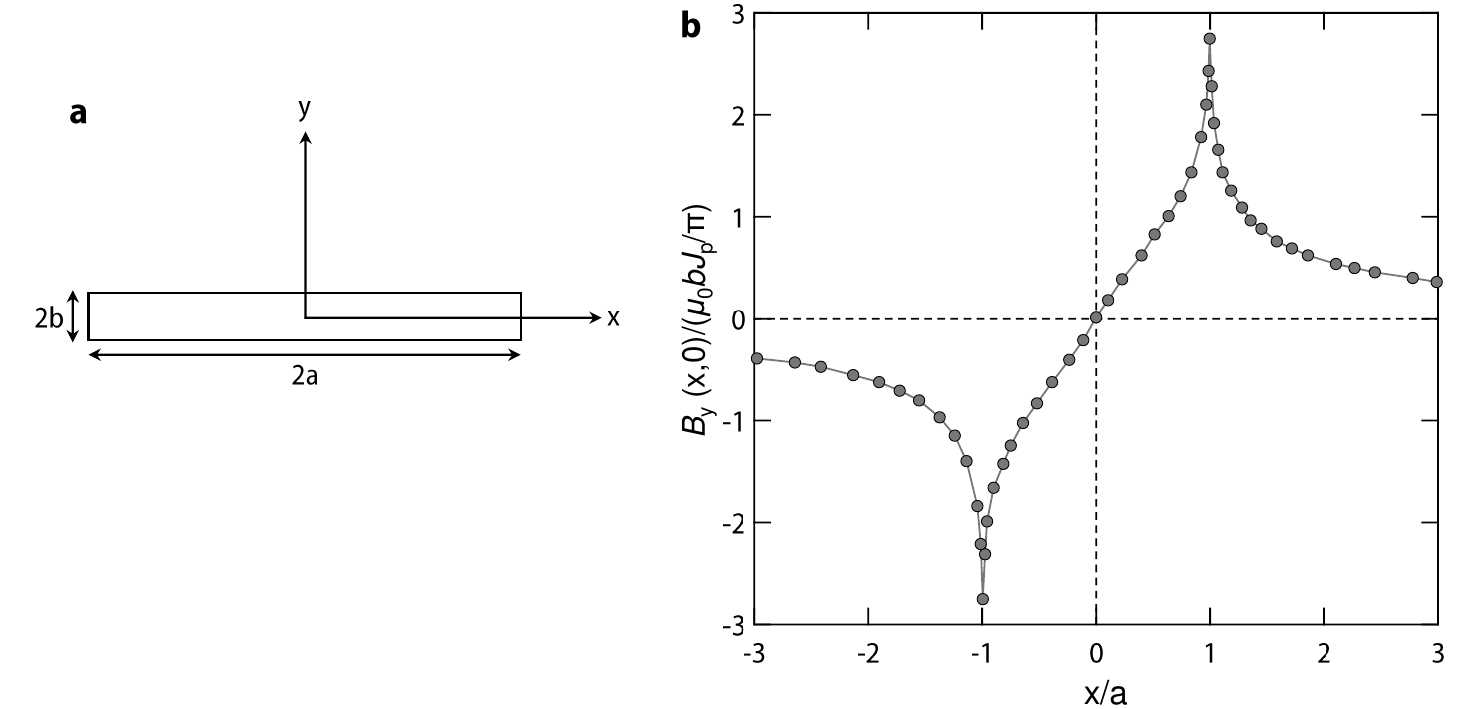}
		\caption{ \textbf{a.} Geometry of the superconducting strip where width of the strip is 2a and the thickness is 2b. In this geometry current is sent along the z-axis which is perpendicular to x and y axes.\textbf{b.} Out of plane component of the magnetic field along the x-axis for y=0. This plot has been reproduced with permission by digitising the 3$^{\textnormal{rd}}$ figure of the reference~\cite{BabaeiBrojeny:2005p888}. }
		\label{SFig12}}
\end{figure}

\clearpage

\noindent\textbf{Supplementary Note 7: Effect of residual field in the magnet}

While we have followed the standard protocol to bring down the magnetic field to zero by oscillating the field, there is always a possibility of very small residual field during the measurement. In order to completely get rid of residual field in our system, we have purposely warmed our superconducting magnet (Oxford Instruments) to room temperature and we waited at room temperature for two weeks to make sure complete evaporations of all cryogens (liquid helium, liquid nitrogen) from the system. This makes sure that there is no residual field present in the magnet. Thereafter the system was freshly precooled, and  $I$-$V$ measurements were performed without applying any magnetic field. In Supplementary Figure \ref{SFig13} we show the comparison of $I$-$V$ data recorded at one representative temperature (1.46 K) for 14 nm AlO$_\textnormal{x}$ film on KTaO$_3$ (111) ($I$ along [11$\bar{2}$]) before and after warming the magnet. As evident curves look same, signifying that residual field in the measurement setup has no role for our observation. 

\begin{figure}[h]
	\centering{
		\includegraphics[scale=0.55]{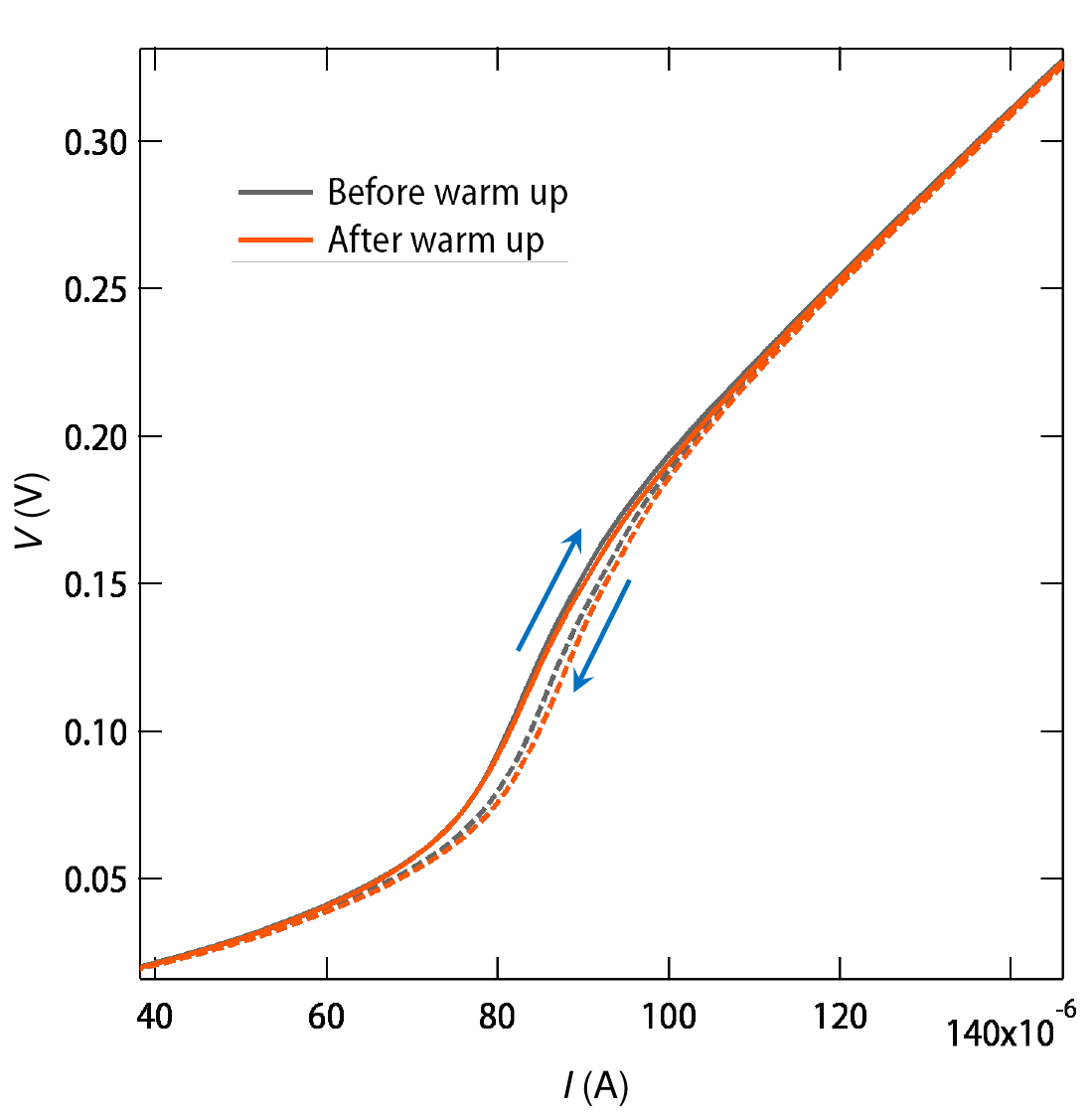}
		\caption{ $I$-$V$ characteristics recorded at 1.46 K for 14 nm AlO$_\textnormal{x}$/KTaO$_3$ (111) sample ($I$ along [11$\bar{2}$]) before and after warming the magnet.}
		\label{SFig13}}
\end{figure}

\clearpage

\noindent\textbf{Supplementary Note 8: Vortex velocity calculation in absence of magnetic field.}

According to  the Gor’kov–Josephson relation~\cite{josephson:1965p419,halperin:1979p599}, the expression for voltage drop is given by

\begin{equation}
V= n_\textnormal{f}\Phi_0vL   \label{seq:3}
\end{equation}

\noindent where $n_\textnormal{f}$ is the free vortex density, $\Phi$$_0$ is the magnetic flux quantum and $L$ is the distance between voltage probes. This expression is markedly distinct from the bulk type II superconductors where $V$ under perpendicular magnetic field is given by $V$=$v$$B$$L$. This basic difference emerges from the fact that in bulk type II superconductors the $n_\textnormal{f}$ is decided by just the magnitude of $B$ and hence $V$ is independent of temperature and current, which is not the case in 2D superconductors as discussed earlier. In the following we first discuss the steps followed to calculate $n_\textnormal{f}$ at a given temperature and current across BKT transition in absence of $B$:

(i) Case I ($I$$\rightarrow$ 0 limit): At absolute zero current and below $T$$_{\textnormal{BKT}}$, vortex and antivortex always exist as bound pairs and hence the resistance should be zero as there are no free vortices. However, application of how so ever small current always leads to Ohmic dissipation in $I$$\rightarrow$ 0 limit due to breaking of few  weakly bound vortex-antivortex pairs~\cite{Jos:2013p} as discussed in main text. 

Between $T$$_{\textnormal{BKT}}$ $\leq$ $T$ $\leq$$T$$_{\textnormal{C}}$, thermal fluctuation leads to breaking of vortex-antivortex pairs interacting at a distance $r$ larger than the correlation length ($\xi$$_v$) and bound pairs still exist for $r$ $<$ $\xi$$_v$. In order to calculate $n_\textnormal{f}$ in $I$$\rightarrow$ 0 limit we use the following universal relation~\cite{Jos:2013p}

\begin{equation}
R=R_\textnormal{N}2\pi\xi^2n_\textnormal{f}    \label{seq:4}
\end{equation}

\noindent where $R_\textnormal{N}$ is the normal state resistance. For estimating $R$, we fit the $I$-$V$ curve from 0-$\SI{10}{\micro\ampere}$ with Ohm's law where the $I$-$V$ curve is indeed linear. The obtained resistance is then plugged in Supplementary equation \ref{seq:4} along with $R_\textnormal{N}$ at 5 K and $n_\textnormal{f}$ is back calculated. For estimation of $\xi$ we use the relation $\xi$($T$) = $\xi$(0)(1-$T$/$T$$_{\textnormal{C}}$)$^{-1/2}$. We denote the $n_\textnormal{f}$ estimated in $I$$\rightarrow$ 0 limit by $n_\textnormal{f0}$.

(ii) Case II ($I$$^{\alpha}$ regime): This corresponds to non-Ohmic regime where a power law dependence is observed due to breaking of vortex-antivortex bound pairs. This introduces additional vortices at higher currents which is given by $n_\textnormal{fp}$ = $K$$I$$^{\alpha-1}$ where $\alpha$ is the same power factor as discussed in Fig. 3c of the main text and $K$ is the temperature dependent coefficient which is estimated from the following relation~\cite{saito:2020p074003}

\begin{equation}
V=2\pi\xi^2KR_\textnormal{N}I^\alpha   \label{seq:5}
\end{equation}

\noindent by fitting the $I$-$V$ curve in power law regime. Since the voltage instability occurs well beyond the $I$$^{\alpha}$ regime, the total $n_\textnormal{f}$ in absence of $B$ is given by $n_\textnormal{f}$ = $n_{f0}$+$n_\textnormal{fp}$. This value is then put in Supplementary equation \ref{seq:3} and the vortex velocity is back calculated.

\clearpage

\noindent\textbf{Supplementary Note 9: Derivation for temperature dependence of LO instability in BKT system and effect of self-heating.}

Another successful prediction of LO theory has been the temperature dependence of $v^*$ close to $T$$_{\textnormal{C}}$ given by $v^*$ $\propto$ $\Delta$$^{1/2}$ $\propto$ (1-$T$/$T_{\textnormal{C}}$)$^{1/4}$. However, this relation was derived with the constraint that $n_\textnormal{f}$ is independent of temperature which is not the case in 2D superconductors. In the following we first derive the relevant expression of $v^*$ in context of BKT system. For this we follow the approach described in  ref~\cite{klein:1985p413} which predicts the following relation for $\eta$($v$)

\begin{equation}
\eta (v)= \eta(0)-\eta(0) \frac{\delta E}{\Delta}   \label{seq:6}
\end{equation}

\noindent where $\delta$$E$/$\Delta$ is the fraction of quasiparticles that have left the vortex core due to $\delta$$E$ change in the average quasiparticle energy due to electric field from flux-flow. Considering the balance between viscous damping force and Lorentz force along with the Supplementary equation \ref{seq:3} yields $\delta$$E$=$\eta$($v$)$v^2$$n_\textnormal{f}$$\tau$$_\textnormal{e}$/$n_\textnormal{qp}$ where $n_\textnormal{qp}$ is the quasiparticle density in vortex core and $\tau$$_\textnormal{e}$ is the electron's inelastic scattering time. Putting this relation in Supplementary equation \ref{seq:6} , we recover the expression for LO theory given in equation 3 of the main text. Considering the expression $\eta$(0)=$\phi_0$$^2$$d$/2$\pi$$\xi$$^2$$R_\textnormal{N}$ from Bardeen-Stephen law~\cite{saito:2020p074003,Tinkham:2004p} we obtain the expression for $v^*$ as 

\begin{equation}
v^*=\bigg(\frac{2\pi R_\textnormal{N}n_\textnormal{qp}\Delta(T)\xi^2(T)}{\phi_0^2d\tau_\textnormal{e}n_\textnormal{f}(T,B)}\bigg)^{1/2}   \label{seq:7}
\end{equation}

\noindent We note that, this relation is drastically different from the LO theory where the temperature dependence enters only through $\Delta$. Further, considering the Bezuglyj and Shklovskij model for effect of self-heating on FFI near $T_{\textnormal{C}}$ would lead to additional $B$$^{-1/2}$ dependence~\cite{bezuglyj:1992p234,dobrovolskiy:2020p3291}, modifying the Supplementary equation \ref{seq:7} to

\begin{equation}
v^*=k\bigg(\frac{2\pi R_\textnormal{N}n_\textnormal{qp}\Delta(T)\xi^2(T)}{\phi_0^2d\tau_\textnormal{e}n_\textnormal{f}(T,B)}\bigg)^{1/2}B^{-1/2}   \label{seq:8}
\end{equation}

\noindent where $k$ is a constant.

\clearpage

\noindent\textbf{Supplementary Note 10: Calculation of $n_\textnormal{f}$$\xi$$^2$.}

\noindent (i) In the absence of magnetic field, the value $n_\textnormal{f}$$\xi$$^2$ before the voltage instability would be given by $n_\textnormal{f}$$\xi$$^2$ = $n_\textnormal{f0}$$\xi$$^2$+$n_\textnormal{fp}$$\xi$$^2$. $n_\textnormal{f0}$$\xi$$^2$ is calculated using Supplementary equation \ref{seq:4} as discussed earlier and $n_\textnormal{fp}$$\xi$$^2$ is given by $n_\textnormal{fp}$$\xi$$^2$=$K$$\xi$$^2$$I$$^{\alpha-1}$. The product $K$$\xi$$^2$ is estimated from the Supplementary equation \ref{seq:5} and the value of $I$ is taken at the mid of power law region in the $I$-$V$  measurement.

\noindent (ii) Application of $B$ leads to an imbalance between vortex and antivortex by increasing the concentration of one of the species by an amount $n_\textnormal{fB}$~\cite{doniach:p1169}. This leads to an additional resistance which we capture by fitting $I$-$V$ curve (at a fixed $B$) in low current regime ($\SI{0}{\micro\ampere}$ to $\SI{10}{\micro\ampere}$) with Ohm's law. The $n_\textnormal{fB}$$\xi$$^2$ is then calculated using  Supplementary equation \ref{seq:4} where the value of $R_\textnormal{N}$ is taken at 5 K. The total $n_\textnormal{f}$$\xi$$^2$ in presence of magnetic field is then given by $n_\textnormal{f}$$\xi$$^2$= $n_\textnormal{f0}$$\xi$$^2$+$n_\textnormal{fp}$$\xi$$^2$+$n_\textnormal{fB}$$\xi$$^2$.

\clearpage

\noindent\textbf{Supplementary Note 11: Temperature dependence of LO instability across BKT transition}

\begin{figure}[h]
	\centering{
		\includegraphics[scale=0.55]{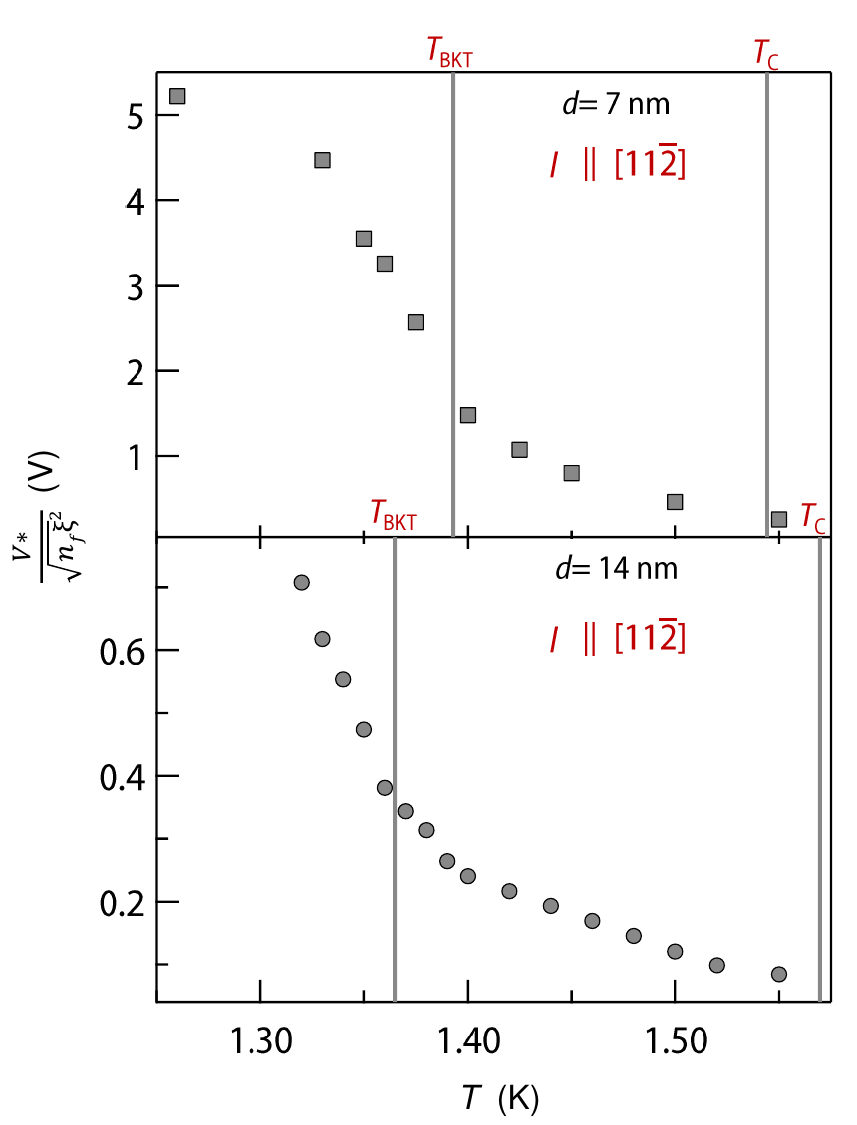}
		\caption{Temperature evolution
			of normalized critical voltage calculated for Hall bar along [11$\bar{2}$] for 7 nm and 14 nm AlO$_\textnormal{x}$/KTaO$_3$ (111) samples. }
		\label{SFig14}}
\end{figure}

The temperature dependence of normalized $V^*$ ($V^*$/$\sqrt{n_\textnormal{f}\xi^2}$) for two samples on Hall bars along [11$\bar{2}$] has been shown in Supplementary Figure \ref{SFig14}. As evident, the curves do not follow the (1-$T$/$T_{\textnormal{C}}$)$^{1/4}$ dependence considering $\Delta$ $\propto$ (1-$T$/$T_{\textnormal{C}}$)$^{1/2}$. Further, we also observe an appreciable enhancement in the normalized $V^*$ below $T$$_{\textnormal{BKT}}$. We propose that such a peculiar behavior could arise due to the dominant role of hot-spot effect below  $T$$_{\textnormal{BKT}}$~\cite{ovadyahu:1980p375} which would lead to overestimation of $V^*$, whereas LO instability would be more applicable near $T$$_{\textnormal{C}}$~\cite{Larkin1975:960}. We note that a similar inference was drawn earlier on YBa$_2$Cu$_3$O$_{7-\delta}$ films where  LO instability was observed along with hot-spots~\cite{xiao:1998p11185}.

\clearpage

\begin{figure}[h]
	\centering{
		\includegraphics[scale=0.55]{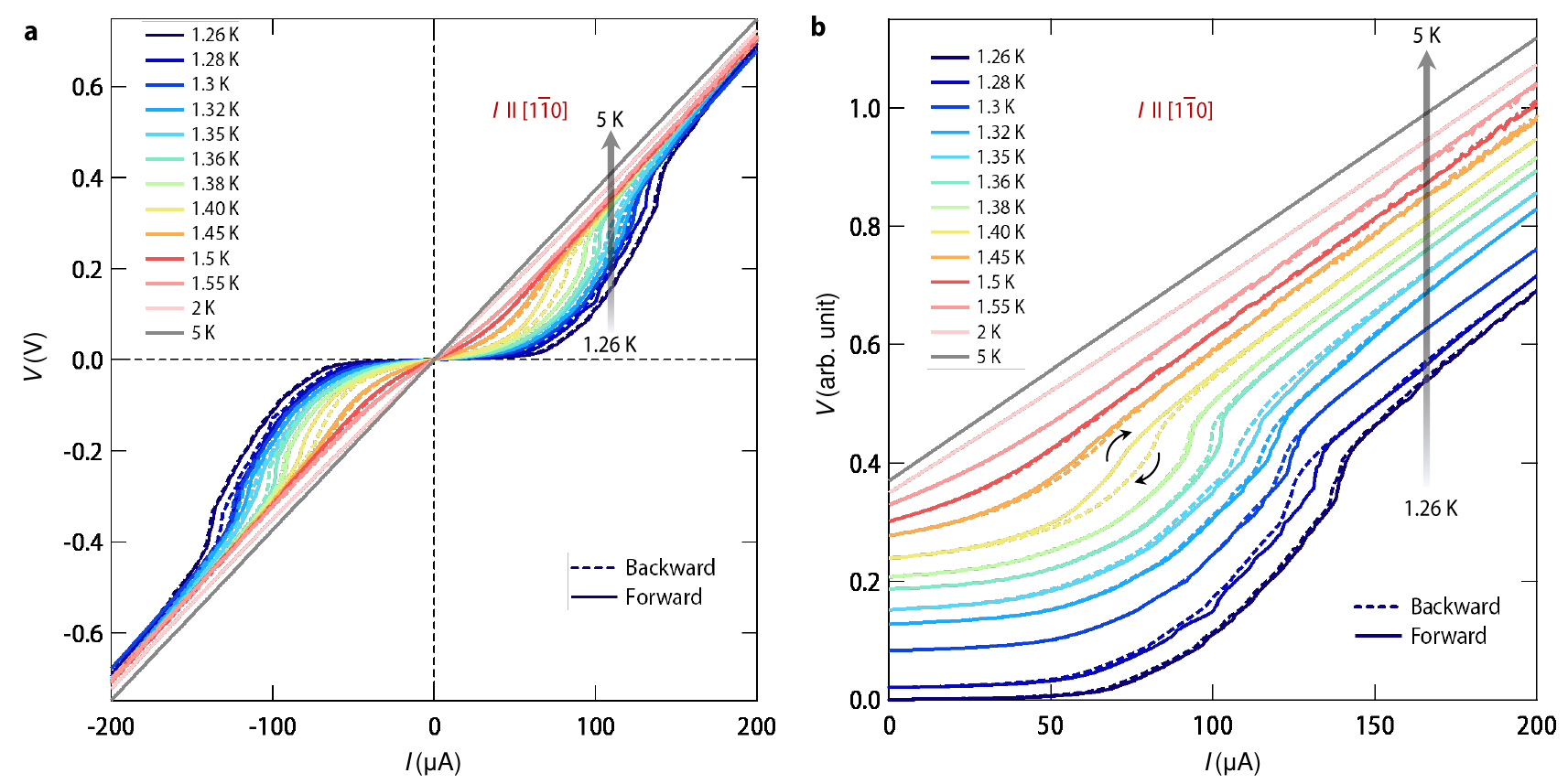}
		\caption{\textbf{a.} Temperature dependent full cycle  $I$-$V$ curves measured in current bias mode for the Hall bar along [1$\bar{1}$0] on 7 nm AlO$_\textnormal{x}$/KTaO$_3$ (111) sample.  For the sake of clarity, $\SI{0}{\micro\ampere}$$\rightarrow$$\SI{200}{\micro\ampere}$ and $\SI{200}{\micro\ampere}$$\rightarrow$$\SI{0}{\micro\ampere}$ branches have not been shown in the plot. Figure \textbf{b} shows the same plot with all the curves (except at 1.26 K) shifted vertically for visual clarity. Similar to $I$ along [11$\bar{2}$], clockwise hysteresis appears above a certain temperature.}
		\label{SFig15}}
\end{figure}

\clearpage

\begin{figure}[h]
	\centering{
		\includegraphics[scale=0.6]{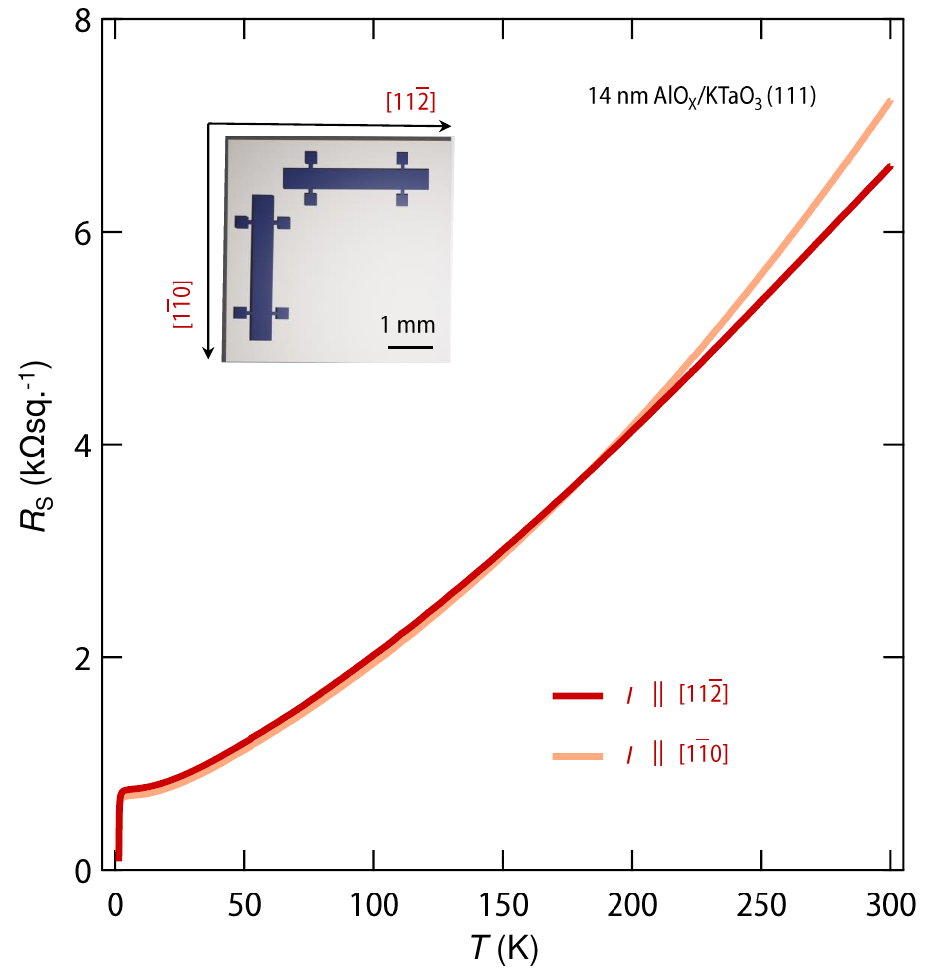}
		\caption{Temperature dependent $R_\textnormal{S}$ of another sample with 14 nm AlO$_\textnormal{x}$ thickness. Similar to 7 nm AlO$_\textnormal{x}$/KTaO$_3$ (111) sample, a little anisotropy is observed between Hall bars made along [11$\bar{2}$] and [1$\bar{1}$0]. $T$$_{\textnormal{C}}$ is found to be 1.57 K and 1.51 K for current along  [11$\bar{2}$] and [1$\bar{1}$0], respectively.}
		\label{SFig16}}
\end{figure}
\clearpage

\begin{figure}[h]
	\centering{
		\includegraphics[scale=0.5]{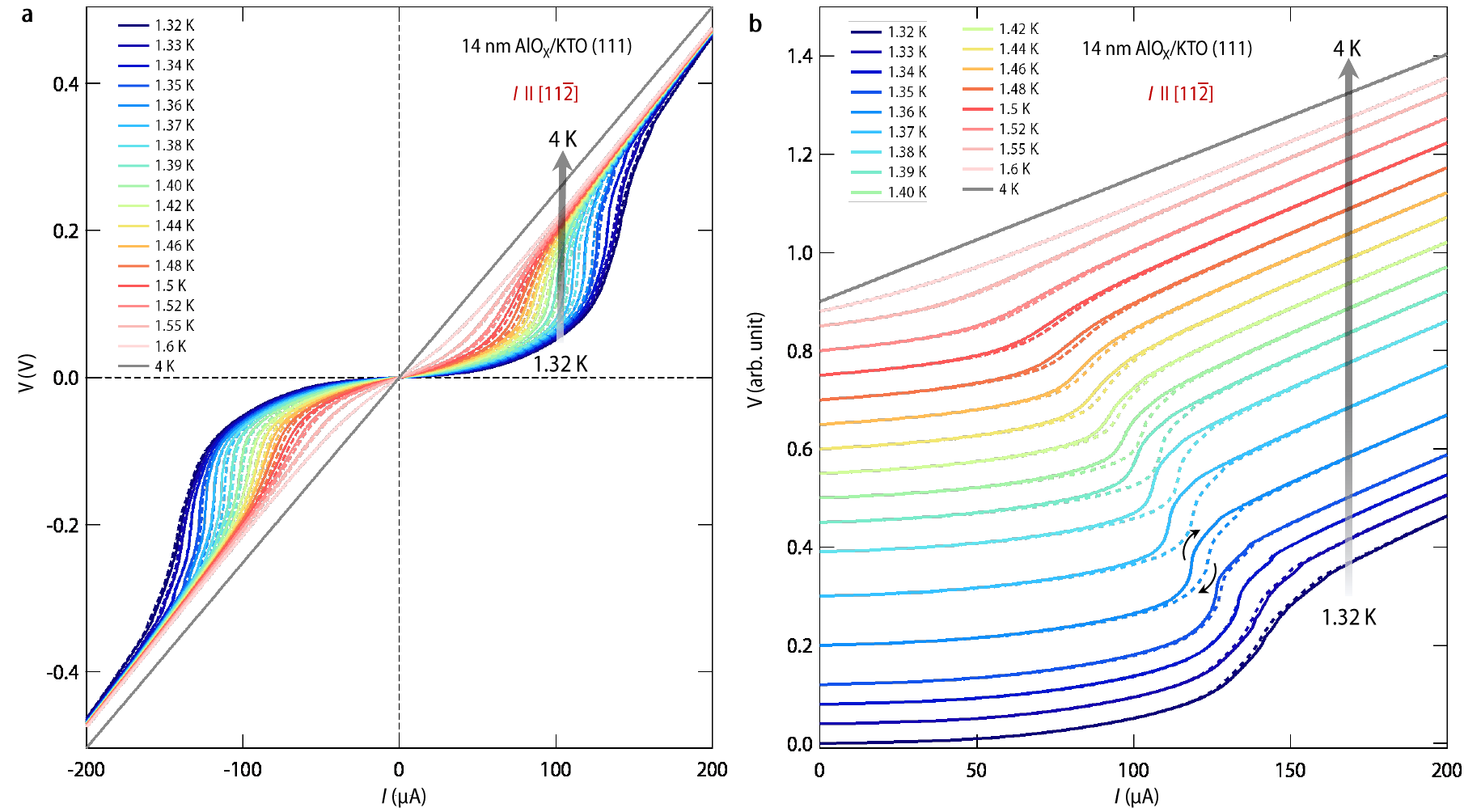}
		\caption{\textbf{a} Temperature dependent full cycle  $I$-$V$ curves measured in current bias mode ($I$ $\parallel$ [11$\bar{2}$]) on another sample with 14 nm AlO$_\textnormal{x}$ thickness.  For the sake of clarity, $\SI{0}{\micro\ampere}$$\rightarrow$$\SI{200}{\micro\ampere}$ and $\SI{200}{\micro\ampere}$$\rightarrow$$\SI{0}{\micro\ampere}$ branches have not been shown in the plot. Figure \textbf{b} shows the same plot with all the curves (except at 1.26 K) shifted vertically for visual clarity. Similar to 7 nm AlO$_\textnormal{x}$/KTaO$_3$ (111) sample, clockwise hysteresis appears above $T_\mathrm{BKT}$.}
		\label{SFig17}}
\end{figure}

\clearpage

\begin{figure}[h]
	\centering{
		\includegraphics[scale=0.53]{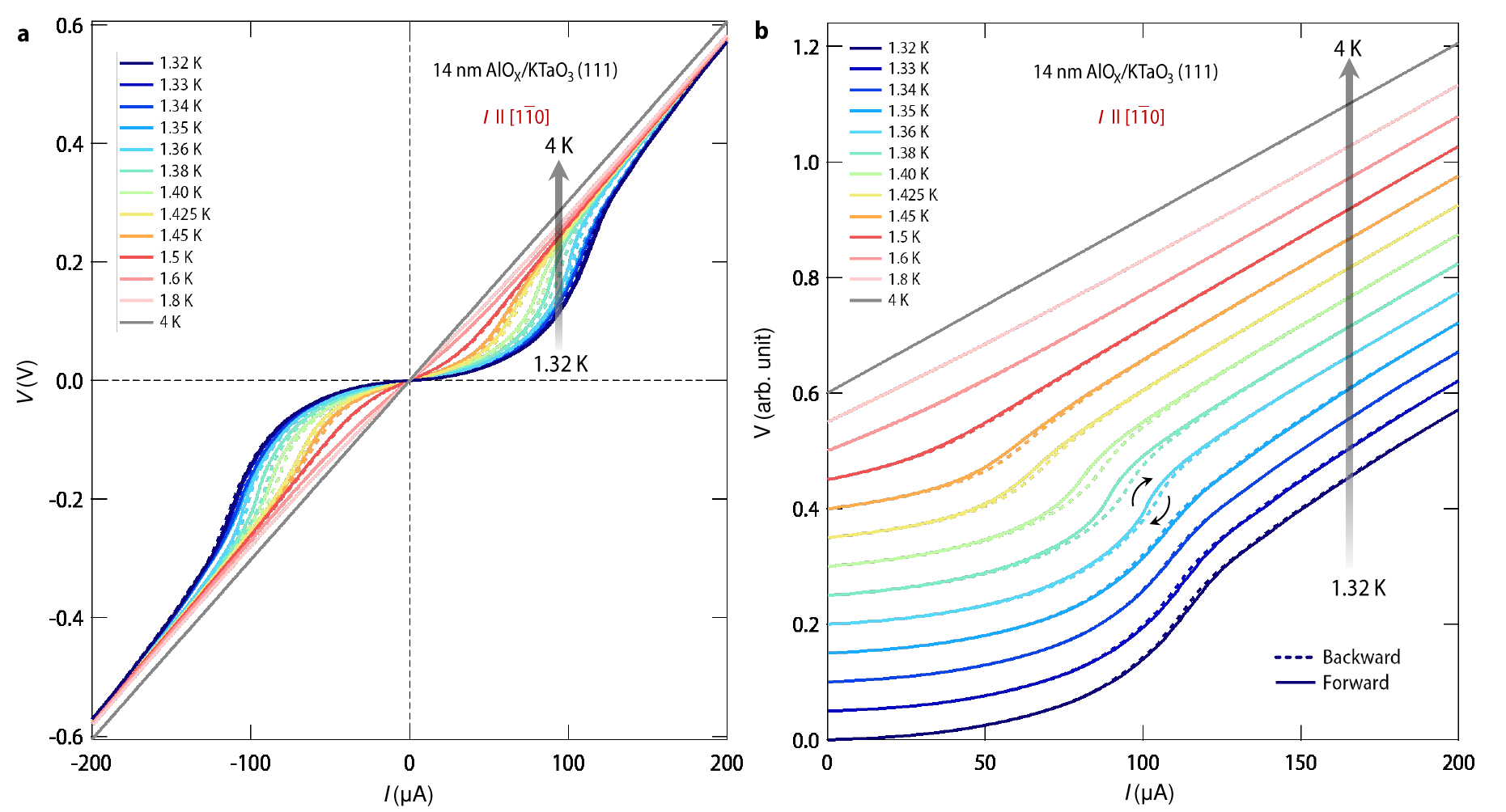}
		\caption{ \textbf{a.} Temperature dependent full cycle  $I$-$V$ curves measured in current bias mode for the Hall bar along [1$\bar{1}$0] on another sample with 14 nm AlO$_\textnormal{x}$ thickness.  For the sake of clarity, $\SI{0}{\micro\ampere}$$\rightarrow$$\SI{200}{\micro\ampere}$ and $\SI{200}{\micro\ampere}$$\rightarrow$$\SI{0}{\micro\ampere}$ branches have not been shown in the plot. Fig. \textbf{b} shows the same plot with all the curves (except at 1.26 K) shifted vertically for visual clarity. Similar to $I$ along [11$\bar{2}$], clockwise hysteresis appears above $T_\mathrm{BKT}$.}
		\label{SFig18}}
\end{figure}

\clearpage

\noindent\textbf{Supplementary Note 12:} Thermally activated flux flow above BKT transition.

\begin{figure}[h]
	\centering{
		\includegraphics[scale=0.45]{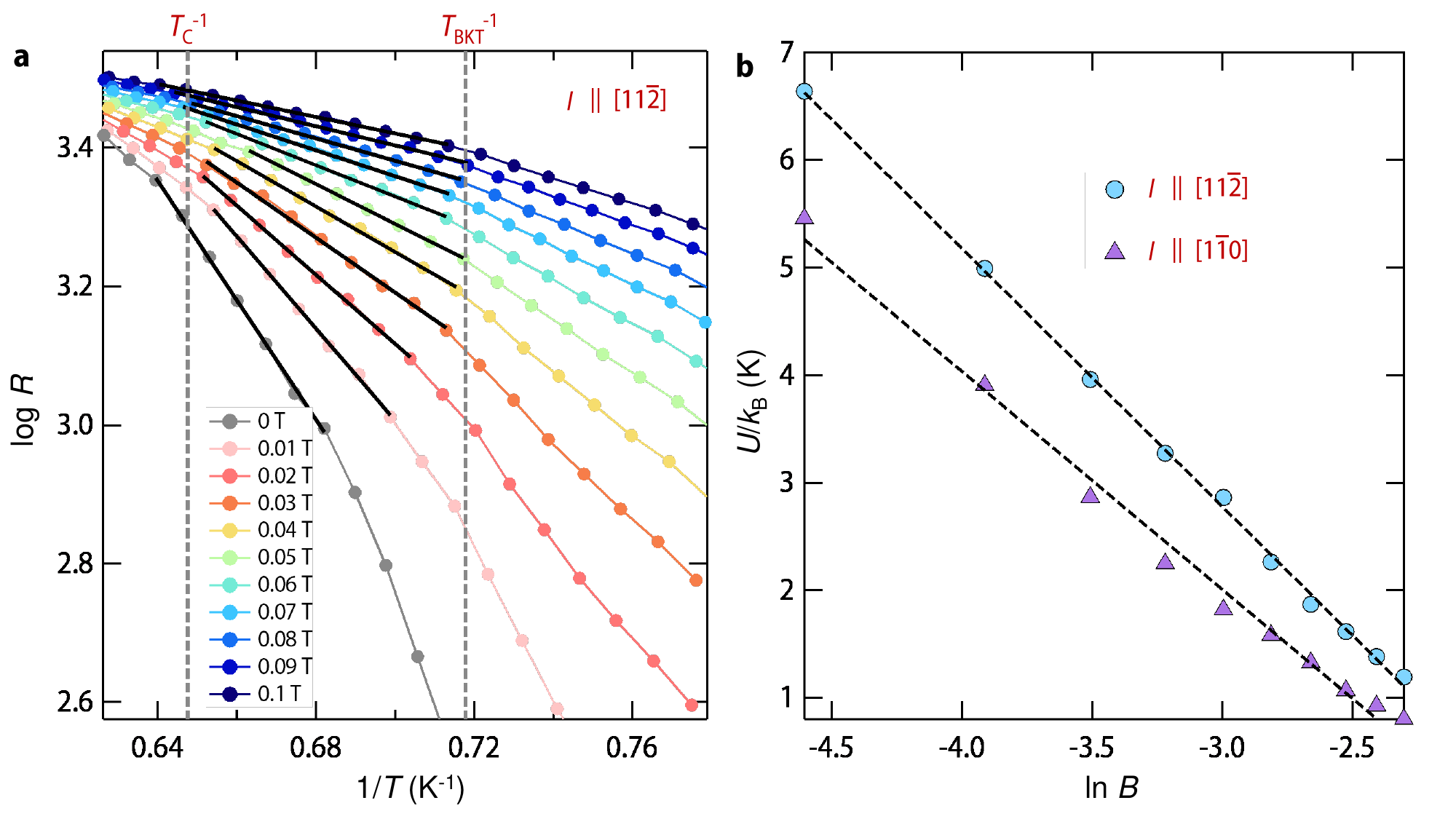}
		\caption{\textbf{a.} Arrhenius plot of resistance at several fixed $B$ for 7 nm AlO$_\textnormal{x}$/KTaO$_3$ sample (for Hall bar along [11$\bar{2}$]). Solid black lines denote fitting with straight line. A similar behavior was observed for the Hall bar along [1$\bar{1}$0] (data not shown). \textbf{a.} Magnetic field dependent energy barrier ($U$) extracted from the linear fit to the activated region for both the Hall bars.}
		\label{SFig19}}
\end{figure}

As discussed in the main text, there is an abundance of free vortices above $T$$_{\textnormal{BKT}}$ which leads to the dissipation in presence of current due to the flux flow. This phenomena happens over an energy barrier ($U$) and has activated behavior leading to the Arrhenius behavior ($R$$\propto$exp(-$U$/$k_\text{B}$$T$)) in the temperature range $T$$_{\textnormal{BKT}}$ and $T$$_{\textnormal{C}}$ as shown in the Supplementary Figure \ref{SFig19}a. The extracted $U$ from the fitting follows the expected dependence on the magnetic field $U$ $\propto$ ln$B$ (see Supplementary Figure \ref{SFig19}b)~\cite{Tsen:2016p208}.



	\end{document}